\documentclass[amsmath,amssymb,twocolumn,aps,pre,twocolumn,superscriptaddress,english]{revtex4-1}  

\pdfoutput=1

\usepackage{lmodern}
\usepackage[T1]{fontenc}
\usepackage[latin9]{inputenc}
\usepackage{amsmath}
\usepackage{amssymb}
\usepackage{mathrsfs}
\usepackage{bm}
\usepackage{esint}
\usepackage{xcolor}
\usepackage{babel}
\usepackage{float}
\usepackage{datetime}
\usepackage{dcolumn}   
\usepackage{graphicx}  
\usepackage{placeins}
\usepackage{hyperref}
\usepackage{float}

\graphicspath{figures/}

\hypersetup{colorlinks,bookmarksopen,bookmarksnumbered,
citecolor=blue,
linkcolor=black,
pdfstartview=green,
urlcolor=purple}

\newcommand{\D}{{\rm d}}

\newcommand{\nn}{\nonumber}

\newcommand{\etak}{\eta_k}
\newcommand{\etakb}{\eta_{\bar{k}}}

\newcommand{\re}{\,\mathrm{Re}\,}
\newcommand{\im}{\,\mathrm{Im}\,}

\newcommand{\Eqref}[1]{Eq.~\eqref{#1}}
\newcommand{\fref}[1]{Fig.~\ref{#1}}

\graphicspath{{figures/}}

\def\be{\begin{equation}}
\def\ee{\end{equation}}
\def\bea{\begin{eqnarray}}
\def\eea{\end{eqnarray}}

\begin{document}

\title{Interface geometry of binary mixtures on curved substrates}

\date{\today}

\author{Piermarco Fonda}
\email{fonda@lorentz.leidenuniv.nl}
\affiliation{Instituut-Lorentz, Universiteit Leiden, P.O. Box 9506, 2300 RA Leiden, The Netherlands}
\author{Melissa Rinaldin}
\affiliation{Instituut-Lorentz, Universiteit Leiden, P.O. Box 9506, 2300 RA Leiden, The Netherlands}
\affiliation{Huygens-Kamerlingh Onnes Lab, Universiteit Leiden, P. O. Box 9504, 2300 RA Leiden, The Netherlands}
\author{Daniela J. Kraft}
\affiliation{Huygens-Kamerlingh Onnes Lab, Universiteit Leiden, P. O. Box 9504, 2300 RA Leiden, The Netherlands}
\author{Luca Giomi}
\email{giomi@lorentz.leidenuniv.nl}
\affiliation{Instituut-Lorentz, Universiteit Leiden, P.O. Box 9506, 2300 RA Leiden, The Netherlands}

\begin{abstract}
Motivated by recent experimental work on multicomponent lipid membranes supported by colloidal scaffolds, we report an exhaustive theoretical investigation of the equilibrium configurations of binary mixtures on curved substrates. Starting from the J\"ulicher-Lipowsky generalization of the Canham-Helfrich free energy to multicomponent membranes, we derive a number of exact relations governing the structure of an interface separating two lipid phases on arbitrarily shaped substrates and its stability. We then restrict our analysis to four classes of surfaces of both applied and conceptual interest: the sphere, axisymmetric surfaces, minimal surfaces and developable surfaces. For each class we investigate how the structure of the geometry and topology of the interface is affected by the shape of the substrate and we make various testable predictions. Our work sheds light on the subtle interaction mechanism between membrane shape and its chemical composition and provides a solid framework for interpreting results from experiments on supported lipid bilayers.
\end{abstract}

\maketitle

\section{Introduction}
\label{sec:intro}

Lipid bilayers are ubiquitous in living systems and have been firmly established as the universal basis for cell-membrane structure \cite{alberts2014molecular}. They protect the interior of the cell from the environment, enclose internal organelles and mediate all the interactions between the various compartments of the cell. Inevitably, high structural complexity is required to accomplish the enormous variety of tasks the cell needs to perform, as it is demonstrated by the myriad of specialized molecules and molecular complexes comprising cellular membranes. 

\textit{In vivo}, membrane heterogeneity is believed to be obtained through the formation of specialized domains \cite{Sezgin2017}. The physical and chemical mechanisms behind the formation and the stability of these domains have been debated in the literature for a long time \cite{Sevcsik2016}. Despite lack of general consensus, experimental evidence from artificial membranes indicates that thermodynamic stability is, at least partially, involved in the process \cite{Rosetti2017}. 
\textit{Artificial} model lipid bilayers are often obtained from self-assembled ternary mixtures of saturated and unsaturated lipids which, under the right external conditions, spontaneously phase-separate and equilibrate towards a state of liquid/liquid phase coexistence. The two phases have different internal order and are labelled as liquid ordered (LO) and liquid disordered (LD) \cite{Ipsen1987}. Various physical properties of these phases, such as thickness and mobility, influence and are influenced by the local membrane shape. Even though the connection with biological membranes remains open to debate \cite{Rosetti2017}, artificial membranes surely are a useful tool to understand one of the fundamental building blocks of life.

Phase separation in artificial lipid bilayers has been investigated for over four decades \cite{Wu1975,Gebhardt1977} and the interplay between membrane shape, domain formation and lateral displacement has been studied in several experimental set-ups \cite{Baumgart2003,Roux2005,parthasarathy2006curvature,Pencer2008,Sorre2009,Ursell2009,Rinaldin2018}. The coexistence of two-dimensional phases implies that a stable linear interface must exist, dividing the membrane into different domains. As in every phase coexistence, this interface has a non-vanishing line tension \cite{Benvegnu1992}.

Alongside experiments, comparable effort has been done on the theoretical side, with the goal of constructing models able to account for the experimental observation. The various approaches can be roughly divided into two main classes. The first one, pioneered by the works of Leibler and Andelman \cite{Leibler1986,Leibler1987}, focuses on the statistical nature of phase separation, treating the membrane as a set of concentration fields interacting with the environment. These fields and their associated thermodynamic potentials are, ideally, emergent mean-field descriptions of the underlying coarse molecular structure. In contrast, the second approach is \textit{geometrical} and treats lipid domains as regions on a two-dimensional surface bounded by one-dimensional interfaces. This view falls within the fluid-mosaic model \cite{Singer1972} and is a natural generalization of the Canham-Helfrich approach \cite{Canham1970,Helfrich1973} to multi-component membranes, first introduced by J\"ulicher, Lipowsky and Seifert in \cite{Seifert1993,Julicher1993,Lipowsky1993a}. 

Here we follow the latter geometric approach and model phase domains as perfectly thin two-dimensional surfaces. Motivated by recent experiments on scaffolded lipid vesicles (i.e. lipid vesicles internally supported by a colloidal particle \cite{Rinaldin2018}), we restrict our analysis to the case of membranes with fixed geometry, such that the only degree of freedom of the system is the position of the interface: the free energy is  a functional of embedded curves. This assumption is appropriate for membranes which are attached to some support and are not free to change their shape.

The central focus of this work is the shape of the interface and how it is influenced by the underling geometry of the membrane. Interfacial lines are obtained as solutions of an interface equation and need to be stable against fluctuations. These requirements are significantly more intricate than interface problems in homogeneous and isotropic environments. For instance, coexisting phases in three-dimensional Euclidean space tend to minimize their contact area and the resulting interface is either planar or spherical, in case of non-zero Laplace pressure. Similarly, on a two-dimensional flat plane, interfaces are either straight lines or circles. 

As we will demonstrate in the remaining sections, the scenario changes dramatically for non-flat membranes. Spatial curvature introduces three essential features that are not present on flat substrates. First, curves on surfaces can be simultaneously curved and length-minimizing (i.e. geodesic). As a consequence, stable closed interfaces can exist on a curved substrate even for vanishing Laplace pressure. Second, as different lipid phases have, generally, different elastic moduli (with the LD phase being more compliant to bending than the LO phase), non-uniform substrate curvature can drive the segregation of lipid domains, with the stiffer phase preferentially located in regions of low curvature at the expenses of the softer phase (i.e. geometric pinning). Third, the surface curvature directly influences the stability of interfaces. In particular, interfaces located in regions of negative Gaussian curvature (i.e. saddle-like) generally tend to be more stable, as any deviation from their original shape inevitably produces an increase in length. 

We stress that, although lipid membranes represent our main inspiration, we study the more general problem of interfacial equilibrium when the ambient curvature influences the energy landscape: our results, therefore, apply to any two-dimensional system with coexisting phases. A non-exhaustive list of additional theoretical works on coexisting fluid domains, separated by a one-dimensional interface, is given in Refs. \cite{Seul1995,Julicher1996,Baumgart2005,Elliott2010,envirobias}. Most of these works, however, focus on lipid vesicles, where both the shape of the membrane and the structure of the phase domains is free to vary. This problem is generally harder than the one addressed here and often analytically intractable. In a few special cases, such as that of axisymmetric surfaces, some progress can be made \cite{Julicher1996,Baumgart2005,Das2009}, under the non-necessary assumption that also the interface inherits the rotational symmetry of the substrate. While keeping the membrane geometry fixed, we relieve any restriction on the interface and provide a more general picture.

This paper is organized as follows: in Section \ref{sec:curved} we write down the free energy functional, which depends only on the position of the interface(s) on the membrane. We compute the first and second variational derivative of this functional and find general stability conditions. In Sec. \ref{sec:lines on surfaces} we show how closed interfaces are stabilized by negative Gaussian curvature. In Sec. \ref{sec:local expansion} we show the local effect of curvature on an arbitrary surface. Sec. \ref{sec:examples} is devoted to the study of specific classes of surfaces: we study the sphere (Sec. \ref{sec:spheres}), axisymmetric surfaces (Sec. \ref{sec:axisymmetric}), minimal surfaces (Sec. \ref{sec:minimal}) and developable surfaces (Sec. \ref{sec:developable}). In Sec. \ref{sec:discussion} we give an overview of the results and discuss future directions. The Appendices are dedicated to the mathematical details of the results. In Appendix \ref{app:notation} we review the general theory of embedded curves. In Appendix \ref{app:variational} we show how to compute variational derivatives of geometric functionals and how the topology of the interface influences the energy landscape. In Appendix \ref{app:curves on minimal surfaces} we derive our results on minimal surfaces, including how, via the Weierstrass-Enneper representation, we can map the interface equation into an equation on the complex plane. In Appendix \ref{app:curves on developable surfaces} we derive our results on developable surfaces and explain the analogy with closed orbits of charged particles in spatially varying magnetic fields.

\section{Interfaces in multicomponent vesicles}

\label{sec:curved}

\begin{figure}
\includegraphics[width=\columnwidth]{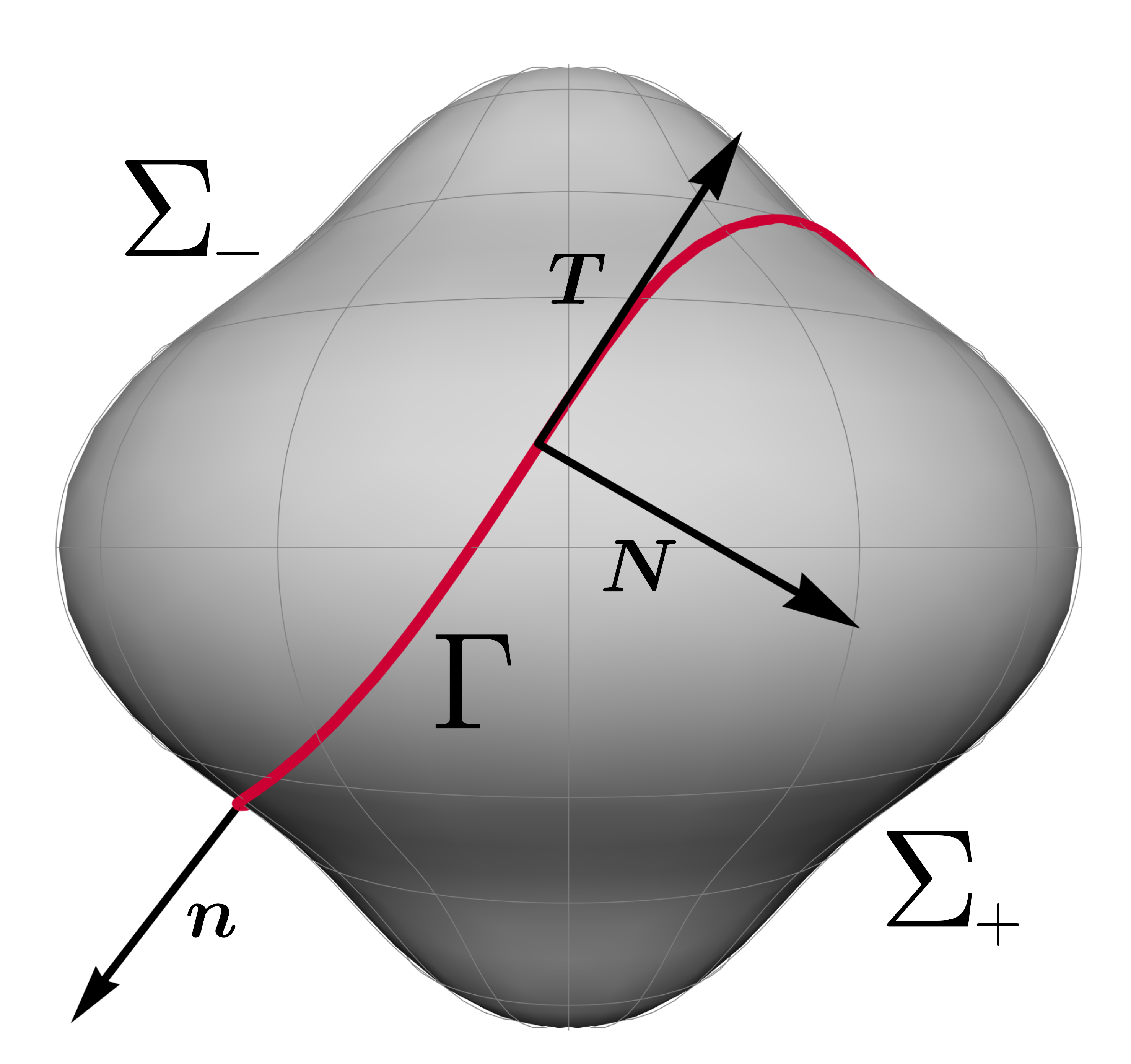}
\caption{The surface $\Sigma$ is partitioned into (multiple) connected domains $\Sigma_+$ and $\Sigma_-$, separated by the linear interface $\Gamma$. The tangent and normal two-vectors of the curve are $T^i$ and $N^i$, which together form a local basis for the tangent space of $\Sigma$. In the figure we show their three-dimensional representation $\bm{T}$ and $\bm{N}$, along with the normal to the surface $\bm{n}$. The three orthonormal vectors $\{ \bm{T}, \bm{N}, \bm{n}\}$ form the \textit{Darboux frame} (or material frame) of $\Gamma$.}
\label{fig:figure 1}
\end{figure}

Following the classic approach introduced by Canham and Helfrich \cite{Canham1970,Helfrich1973}, we model a lipid vesicle as a closed surface $\Sigma$, whose free energy is expressed in terms of geometrically invariant combinations of the metric tensor $g_{ij}$ and the extrinsic curvature tensor $K_{ij}$. Some basic properties of these objects are reviewed in Appendix A1. In the presence of multiple lipid phases, here labelled by ``$+$'' and ``$-$'', the Canham-Helfrich free energy can be generalized as follows:
\be
F
=
\sum_{\alpha=\pm}
\int_{\Sigma_\alpha} {\rm d}A \,
\left( 
\lambda_\alpha
+
k_\alpha H^2
+
\bar{k}_\alpha
K
\right)
+
\sigma
\int_\Gamma {\rm d}s 
\,,
\label{freeenergy}
\ee
where $\Sigma_{+}$ and $\Sigma_{-}$ represent the portions of the surface occupied by the ``$+$'' and ``$-$'' phase respectively and $H$ and $K$ denotes the surface mean and Gaussian curvature. These regions are not necessarily simply connected and might comprise multiple disconnected domains. $\Gamma$ denotes the interface between the two lipid phases and consists of one or more closed curves over the surface $\Sigma$. The functional \Eqref{freeenergy} was first proposed in Ref. \cite{Julicher1993}.
The coefficients $k_\alpha$ and $\bar{k}_\alpha$ are known respectively as the bending and Gaussian rigidities, whereas $\sigma$ is the interfacial line tension. Finally, $\lambda_\pm$ are Lagrange multipliers, analogous to chemical potentials or surface tensions, enforcing incompressibility of both lipid phases. They are chosen such that
\be
\int_{\Sigma^{-}}\D A+\int_{\Sigma^{+}}\D A = \varphi A_{\Sigma} + (1-\varphi) A_{\Sigma}\;,
\label{area fractions}
\ee
where $\varphi$ represents the area fraction occupied by the ``$-$'' phase, $1-\varphi$ is the area fraction occupied by the ``$+$'' phase and $A_{\Sigma}=\int_{\Sigma}{\rm d}A$ is the total surface area.
\Eqref{freeenergy} can be generalized by adding a spontaneous curvature term, but this is neglected here under the assumption that the two leaflets forming the lipid bilayer have identical geometry and chemical composition. 

Minimizing \Eqref{freeenergy} is, in general, a formidable task as the Euler-Lagrange variations of $F$, with respect to both membrane shape and interface position, are non-linear and mutually coupled (an explicit derivation of these equations using a geometric approach can be found in Ref. \cite{Elliott2010}). As a result of this coupling, the three-dimensional shape of each domain depends non-trivially on the position of the interface and vice-versa (a showcase of possible solutions is given e.g. in Ref. \cite{Wang2008}).

Motivated by recent experimental results on scaffolded lipid vesicles \cite{Rinaldin2018}, we here overcome this complication by assuming the geometry of the membrane to be fixed. Since the shape of $\Sigma$ cannot be changed, the only relevant degree of freedom is the position of the interface $\Gamma$. The problem of finding minima of \Eqref{freeenergy} is thus reduced to the simpler task of finding lines on a fixed two-dimensional surface, provided they satisfy specific geometrical constraints. Physically, this can be achieved if any membrane fluctuation in the direction normal to $\Sigma$ is suppressed. Despite this simplification, the phenomenology arising from this problem is remarkably rich and further provides an avenue to discriminate between the roles of the two bending moduli and how they conspire with the membrane geometry.

By calculating the normal variation of $F$ with respect to the position of the interface, we find the following equilibrium condition (see Appendix \ref{app:first variation} for details):
\be
\sigma \kappa_g
=
\Delta k \left. H^2 \right.
+
\Delta \bar{k} \left. K \right.
+
\Delta \lambda
\,,
\label{interfaceq}
\ee
where $\kappa_g$ is the \textit{signed} geodesic curvature of $\Gamma$ - with the convention that $\kappa_g>0$ for a convex domain of ``$-$'' phase - and the curvatures $H$ and $K$ are calculated along the interface $\Gamma$ (see Appendix A2 for expression of $H$ and $K$ in the material frame of $\Gamma$). We define the difference in bending rigidities of the two phases as $\Delta k = k_+ - k_-$ and $\Delta \bar{k}=\bar{k}_{+}-\bar{k}_{-}$. Furthermore, if $\Delta \lambda \neq 0$, it is intended that the interface is partitioning $\Sigma$ in such a way that the fractional area occupied by a single phase is fixed. 

This seemingly simple equation, which holds for arbitrary surfaces $\Sigma$, might or might not be analytically tractable, depending on the complexity of the underlying surface. It usually admits multiple non-equivalent stable and metastable solutions. 
Calculating the second variation of \Eqref{freeenergy} (see Appendix \ref{app:second variation}) yields the following stability condition of the interface under an arbitrary perturbation: 
\be
\sigma \left( K + \kappa_g^2 \right)
+
\Delta k \; \nabla_{\bm{N}} H^2 
+
\Delta \bar{k} \;  \nabla_{\bm{N}} K
<
0
\,,
\label{stability} 
\ee
where $\nabla_{\bm{N}}$ is the surface-covariant directional derivative along the tangent-normal of $\Gamma$ (the vector $\bm{N}$ in \fref{fig:figure 1}). If the conservation of the area is imposed onto fluctuations, then \Eqref{stability} has to be modified in a non-trivial way. 

To reduce the number of independent parameters in \Eqref{interfaceq}, we introduce the dimensionless numbers:
\be
\etak = \frac{\Delta k}{\sigma L}\;,\qquad
\eta_{\bar{k}} = \frac{\Delta \bar{k}}{\sigma L}\;,
\label{etas}
\ee 
expressing the relative contribution of bending and interfacial tension to the total energy. The quantity $L$ denotes the characteristic length of the system and can be chosen, on a case-by-case basis, depending on the symmetry of the surface. 
These numbers are the only necessary parameters that determine the interface position, if and only if the shape of $\Sigma$ is kept fixed. Conversely, when comparing different shapes one should keep in mind that the geometry enters locally into the problem, thus $\eta_{k}$ and $\eta_{\bar{k}}$ only give a general indication of whether force balance at the interface is dominated by bending or tension, but are not sufficient per se to determine the shape of the interface or to predict whether the there will be only two or multiple domains. 

In the following, we will always take $\etak \geq0$ without loss of generality. Since stiffer phases have greater bending rigidity $k$, we often will call the ``$+$'' domains \textit{hard} (so that they correspond to portions of $\Sigma$ occupied by the LO phase) and the ``$-$'' domains \textit{soft} (i.e. consisting of LD phase). 

\subsection{Geodesic and constant geodesic curvature interfaces}

\label{sec:lines on surfaces}

\Eqref{interfaceq} reduces the physical problem of identifying the interface between two lipid phases to the geometrical problem of finding curves embedded on surfaces whose geodesic curvature depends directly on both intrinsic and extrinsic properties of the immersion. This is in general a challenging task, not only because the membrane geometry influences the local behaviour of the interface, but also because for a curve to be an admissible interface it needs to be closed and simple. These are global properties and need to be considered with care. To make progress, in this and the next subsections we will analyze separately the role of each term in \Eqref{freeenergy} and investigate its physical meaning.

As a starting point, let us assume that the local membrane curvature does not influence the interface position, so that $\etak=\eta_{\bar{k}}=0$. Furthermore let us consider the case in which the total area occupied by the lipid phases is not conserved, hence $\Delta \lambda =0$. In practice, this happens if the membrane is in contact with a lipid reservoir. Then, \Eqref{interfaceq} becomes simply 
\be 
\kappa_g = 0 \,,
\label{geodesic equation}
\ee
telling us that $\Gamma$ is a closed geodesic of $\Sigma$. The latter is a curved-space generalization of the intuitive property of interfaces, which pay a fixed energetic cost per unit length, to minimize their extension (similarly, two-dimensional interfaces at equilibrium are minimal surfaces with $H=0$). 

On a flat substrate, the only solutions of \Eqref{geodesic equation} are straight lines. A compact closed surface, on the other hand, allows for richer structures and in particular it admits simple closed geodesics, i.e.  geodesic lines of finite length which do not self-intersect. For example, on a sphere every great circle has $\kappa_g=0$ and for every point on the surface there are infinitely many simple closed geodesics. However, for less symmetric surfaces this might not necessarily be true. This implies that regions of the surface that do not admit closed geodesics cannot host an interface such as the one obtained under the current assumptions. Nonetheless, it is known that every genus zero surface admits at least three simple closed geodesics \cite{grayson1989shortening}.

\begin{figure}
\centering
\includegraphics[width=\columnwidth]{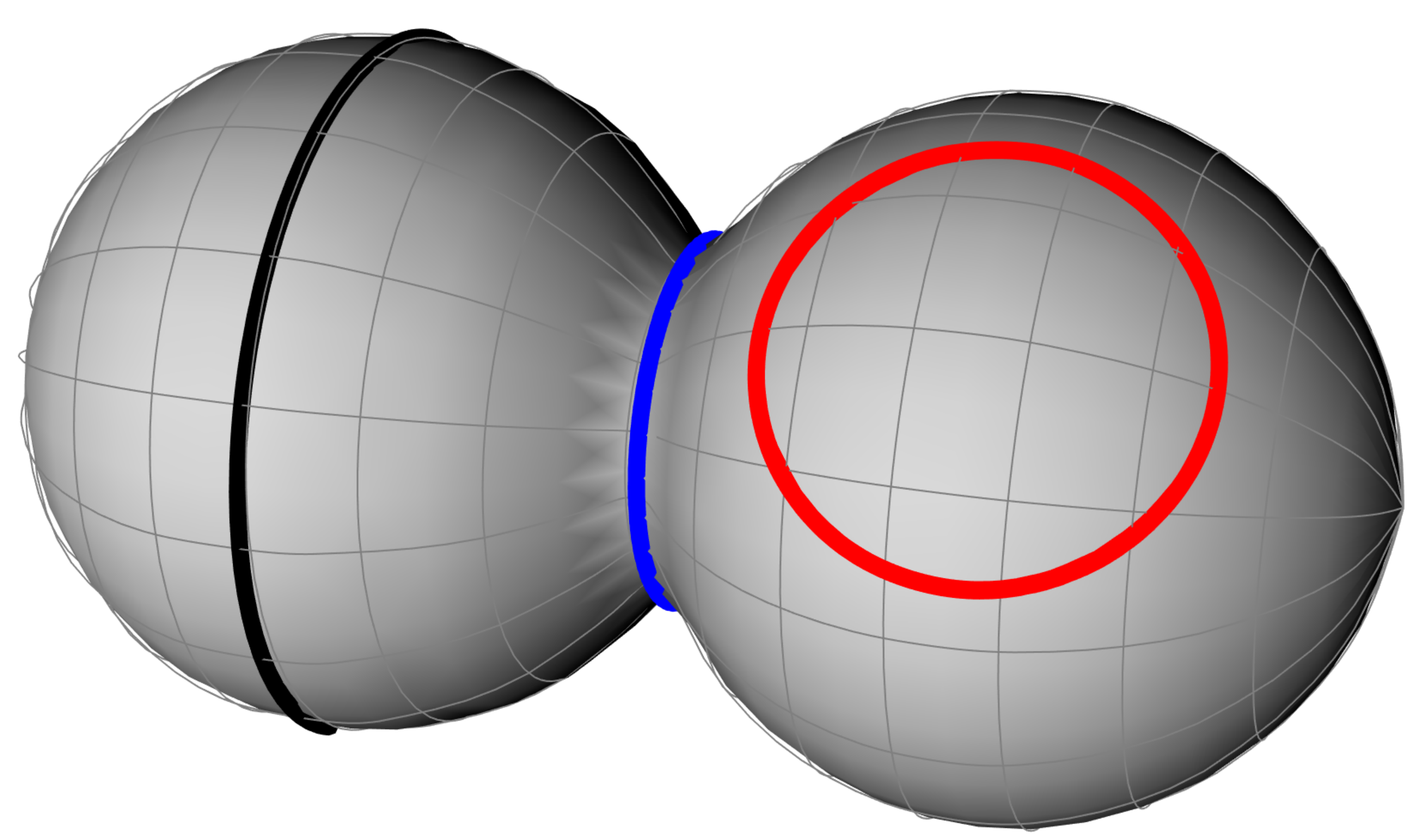}
\caption{
Constant geodesic curvature (CGC) curves on a generic surface. The black line is an unstable closed geodesic: its length can be easily shortened by a shift in any direction. Conversely, the blue line is a stable geodesic, lying along a region of negative $K$, and whose length can only be increased by fluctuations. The red curve is a closed CGC. Since this surface is axisymmetric, meridians and parallels are also principal directions.
This dumbbell-shaped surface was taken from \cite{Rinaldin2018} and is also used to construct the phase diagram of \fref{fig:db phase diagram}.}
\label{fig:curves on db}
\end{figure}

The stability of geodesic interfaces can be easily assessed by taking $\etak = \etakb = 0$ and setting $\kappa_{g}=0$ in \Eqref{stability}. This yields:
\be
K < 0\;,
\label{negative KG}
\ee
thus curves lying in saddle-like regions are inherently stable. This can be intuitively understood by looking at the blue curve in \fref{fig:curves on db}. Moving the interface away from the saddle would inevitably result into an increase of its total length. Conversely, no geodesic lying on regions with positive curvature can represent a stable interface, as its length could always be shortened by a small displacement, as illustrated by the black curve in \fref{fig:curves on db}. In particular, no geodesic of the sphere is stable for non-fixed area fraction $\varphi$.

Next, let us consider the case where the two phases still have identical bending rigidities but their area fractions are kept fixed. \Eqref{interfaceq} yields a curved background analogue of the Young-Laplace equation, namely
\be 
\kappa_g = \frac{\Delta \lambda}{\sigma} \,.
\label{CGC equation}
\ee
Thus, if $\varphi$ is fixed but there is no difference in the elastic moduli, the interface consists of a curve of constant geodesic curvature (CGC), such as the red curve in \fref{fig:curves on db}. We emphasize that $\Delta \lambda$ is determined solely by the area constraint and, if $\Gamma$ consists of multiple disconnected curves, it can take on different values in each of them. This allows the existence of multiple domains bounded by interfaces of constant geodesic curvature (see Appendix \ref{app:gamma topology} for further details). Regardless of their stability, however, configurations featuring multiple domains tend to be metastable as they usually are local minima of the free energy in the absence of a direct coupling with the curvature.

We stress that the stability condition for fixed $\varphi$ is not given by \Eqref{stability}, because only variations that do not change the relative area fractions are allowed, see Appendix \ref{app:second variation}. Unfortunately, the explicit expression of the second variation is not particularly illuminating unless the geometry of $\Sigma$ is made explicit. Therefore we leave further considerations to Section \ref{sec:examples}, where we discuss specific examples.

\subsection{The local effect of curvature}

\label{sec:local expansion}

\begin{figure*}[t]
\includegraphics[width=.9\linewidth]{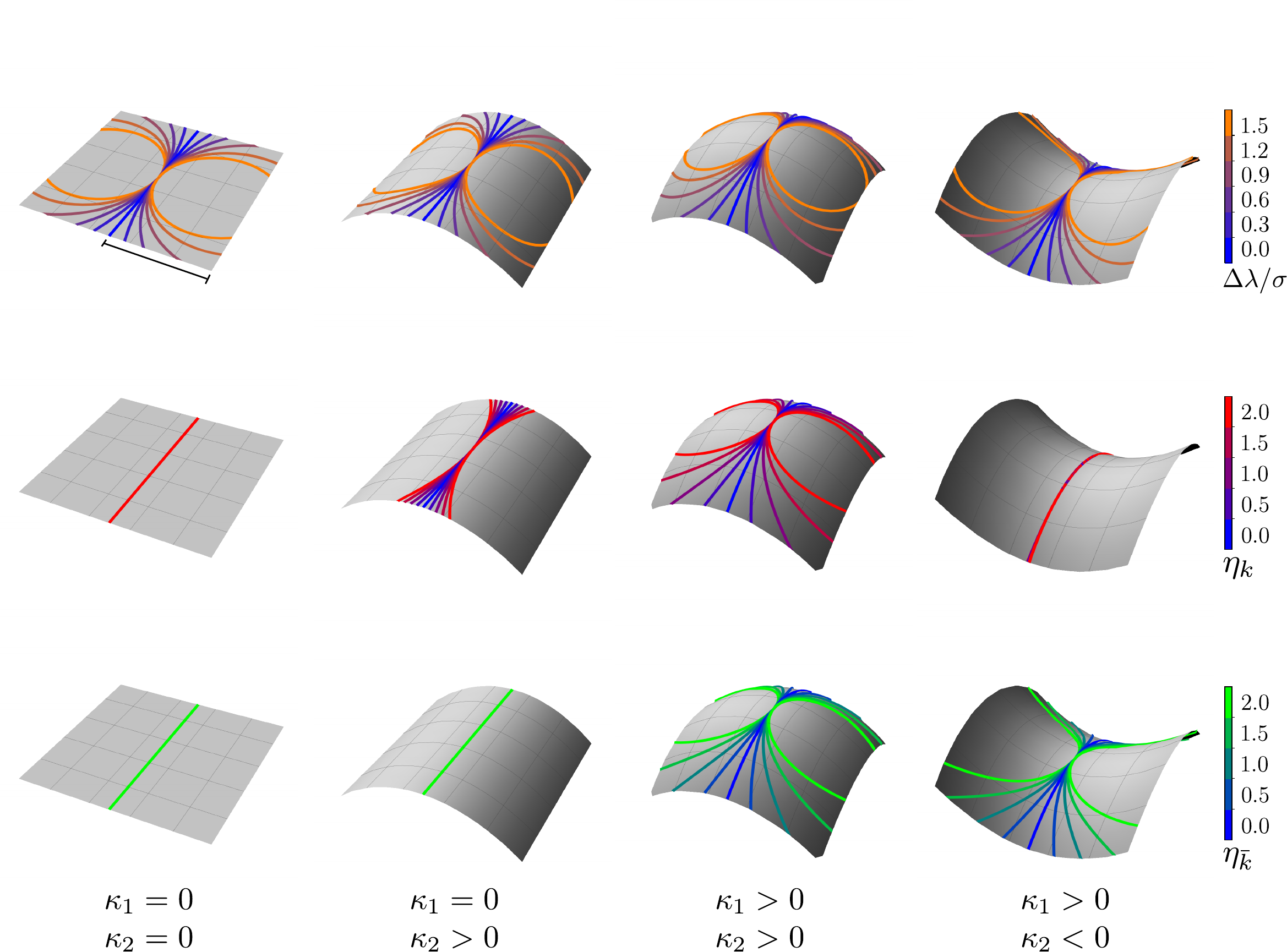}
\caption{
Local effects on the shape of $\Gamma$ of the various terms arising in \Eqref{interfaceq}. The four columns correspond to four different quadric surfaces, parametrized by \Eqref{monge quadric}, with various values of $\kappa_1$ and $\kappa_2$, as shown at the bottom. The first column correspond to a flat plane, the second to a parabolic cylinder, the third to a symmetric paraboloid and the fourth to a hyperboloid. Each row corres.ponds to solutions of \Eqref{interfaceq} where only one of the three terms on the right-hand side is different from zero, shown on the right. Different curve colors correspond to different values of the coupling constants.  Pure blue lines always correspond to geodesic interfaces. Note that the legend on the right refers on the modulus of the couplings, while in the drawing both signs are considered. The scale bar for $L$, used in \Eqref{etas}, is shown in the top left surface. All curves intersect at the point $x=y=0$, at the center of the surface. Notice how $\etak$ does not influence the interface on hyperboloids (it is an almost-minimal surface) and how $\etakb$ does not affect $\Gamma$ on a cylinder, being a developable surface.}
\label{fig:local surfs}
\end{figure*}

In this section, we explore how the local mean and Gaussian curvatures affect the shape of the interface in the presence of inhomogeneous elastic moduli, i.e. $(\etak,\,\etakb) \ne (0,0)$. Any smooth surface can be locally approximated as a quadric, by constructing an adapted Cartesian frame whose origin is a point on the surface, the $x$ and $y$ axes correspond to the principal directions and $z$ to the surface normal $\bm{n}$ (see \fref{fig:figure 1}). 
In a small neighbourhood of the origin, the surface can be approximated with a local Monge patch as
\be
z = \frac{1}{2}(\kappa_1 x^2 + \kappa_2 y^2) \,, 
\label{monge quadric}
\ee
where $\kappa_1$ and $\kappa_2$ are the two principal curvatures at $\{x,y\}=\{0,0\}$. The mean and Gaussian curvature at the origin are $H_0 = \frac{1}{2}(\kappa_1+\kappa_2)$ and $K_0 = \kappa_1 \kappa_2$.
 
An embedded curve can be described with a pair of functions of the arc-length $s$: $\{x,y\}=\{x(s),y(s)\}$. We parametrize the unit tangent along the interface as $\bm{T}=\cos\theta\,\bm{\hat{x}}+\sin\theta\,\bm{\hat{y}}$, where $\bm{\hat{x}}$ and $\bm{\hat{y}}$ are coordinate unit vectors in the $x-$ and $y-$direction and $\theta=\theta(s)$ is the angle between $\bm{T}$ and $\bm{\hat{x}}$. We choose $s$ such that $x(0)=y(0)=0$, and we fix $\theta(0)=\theta_0$ to be the direction of $\bm{T}$ at the origin. Here, and for the rest of the article, we use a dot to indicate differentiation with respect to the arc-length, namely: $\dot{(\cdots)}=\D (\cdots)/ \D s$. Substituting \Eqref{monge quadric} into \Eqref{interfaceq} and expanding for small $s$, we find:
\be
\kappa_g = \dot{\theta}_0 + s \left[\ddot{\theta}_0  -\kappa_{n0} \tau_{g0}\right] + O(s^2) \,,
\label{kg local}
\ee
where $\kappa_n$ and $\tau_g$ are respectively the normal curvature and the geodesic torsion of $\Gamma$ (for definitions, see Appendix \ref{app:curves}). The $0$ subscript denotes the value at the origin. Similarly, we can evaluate and expand up to $O(s^2)$ the surface curvatures along $\Gamma$
\begin{subequations}\label{eq:local_expansion}
\begin{align}
H^2 &= H_0^2 +  H_0 \left[3H_0 K_0 + \kappa_{n0}\left(K_0-6 H_0^2\right) \right]s^2\,,  \label{H2 local} \\[5pt]
K &= K_0 + 2 K_0 \left(K_0-2 H_0 \kappa_{n0} \right)s^2 \,. \label{K local}
\end{align}
\end{subequations}
The lack of linear terms in $s$ in Eqs. \eqref{eq:local_expansion} reflects that the parametrization given in \Eqref{monge quadric} approximates $\Sigma$ at second order in both $x$ and $y$. Substituting Eqs. \eqref{kg local} and \eqref{eq:local_expansion} into \Eqref{interfaceq}, we can solve the resulting equation order by order in powers of $s$.

At order zero we find that \Eqref{interfaceq} constrains the value of $\dot{\theta}_0$. Note that the quantity $r_0=1/\dot{\theta}_0$ is the (signed) radius of curvature of the interface on the tangent plane at $s=0$ (i.e. the radius of the osculating circle on the plane identified by the vectors $\bm{T}$ and $\bm{N}$ of the Darboux frame illustrated in \fref{fig:figure 1}). The interface equation fixes this radius to	
\be
r_{0} = \frac{1}{L(\etak H_{0}^2 + \etakb K_{0}) + \frac{\Delta\lambda}{\sigma}}  \,,
\ee
where $L$ is the length scale used in the definitions \Eqref{etas}. We see that even in the case of non-fixed area fraction, for which we have $\Delta \lambda =0$, the situation is significantly different with respect to the flat case. As a consequence of the substrate local curvature, the interface deviates from a geodesic (for which $r_{0}\rightarrow\infty$), becoming more and more curved the larger is the difference in stiffness between the two lipid phases.

At order $O(s)$ we find the condition $\ddot{\theta}_0 = \kappa_{n0} \tau_{g0}$ which does \textit{not} depend on bending rigidities: it is the same for a geodesic, and states that the rate of change of $r_0$ along $\Gamma$ depends only on the direction of $\bm{T}$. In fact, it vanishes for asymptotic lines (curves with vanishing normal curvature) and for lines of curvature (curves with vanishing geodesic torsion). Higher order contributions are less illuminating, see Appendix \ref{app:local expansion}.

\fref{fig:local surfs} shows the interfaces resulting from a numerical solution of \Eqref{interfaceq} for the quadric surface given by \Eqref{monge quadric}, with different principal curvatures $\kappa_1$, $\kappa_2$ and various $\Delta \lambda$, $\eta_{k}$, $\eta_{\bar{k}}$ values. As expected, while $\Delta \lambda$ has roughly the same effect on $\Gamma$ independently on the surface's curvature (see the first row of \fref{fig:local surfs}), a non-zero curvature coupling produces very different effects depending on the local bending of $\Sigma$.

\section{The effect of curvature for special surfaces}

\label{sec:examples}

The scenario outlined in the previous section applies to arbitrary surfaces. Because of the substrate-dependent nature of the force balance condition expressed by \Eqref{interfaceq} it is not easy to draw general conclusions other than those already discussed. In order to make progress and to develop an intuitive understanding of the \textit{global} effect of the various terms in \Eqref{interfaceq} and of the stability condition of \Eqref{stability}, we will now consider a number of special surfaces, namely spheres (Sec. \ref{sec:spheres}), axisymmetric surfaces (Sec. \ref{sec:axisymmetric}), minimal surfaces (Sec. \ref{sec:minimal}) and developable surfaces (Sec. \ref{sec:developable}). The latter two classes of surfaces are characterized by the property of having vanishing mean and Gaussian curvature respectively, which will allow us to isolate the effect of differences in either bending or Gaussian rigidities. 

\subsection{Spheres}

\label{sec:spheres}

The sphere is the most symmetric closed surface and one of the most common vesicle shapes found in nature, being the absolute minimum of both the area and the bending energy for fixed enclosed volume. All the points of the sphere are umbilic, thus the principal directions of curvature are everywhere undefined. Furthermore, both the mean and the Gaussian curvature are constant throughout the surface and such that $4H^{2}=K=1/R^{2}$, with $R$ the sphere radius.
The total energy of \Eqref{freeenergy} becomes:
\be
F
=
\sum_{\alpha=\pm}
\lambda'_\alpha
\int_{\Sigma_\alpha} {\rm d}A \,
+
\sigma
\int_\Gamma {\rm d}s 
\,,
\label{freeenergy sphere}
\ee
where $\lambda'_\alpha = \lambda_\alpha + (k_\alpha + \bar{k}_\alpha )/R^{2}$ is a constant.  
The interface equation then reduces to \Eqref{CGC equation} with $\Delta\lambda'$ replacing $\Delta\lambda$, independently on the elastic moduli. This corresponds to non-maximal circles of constant geodesic curvature
\be
\kappa_g = \frac{\cot \psi}{R} \,,
\label{CGC sphere}
\ee
where $\psi$ is the usual azimuthal angle in spherical coordinates. The fractional area occupied by such a domain is 
\be
\varphi = \frac{1-\cos \psi}{2} \,.
\ee
Consistent with our convention on the sign of curvatures, we choose $\psi<\pi/2$ for a soft phase domain with $\varphi<1/2$ and $\kappa_g>0$. If the area fractions are not conserved ($\lambda_\alpha=0$), the interface equation admits as solution CGC lines with azimuthal angle:
\be 
\cot \psi = \frac{\etak}{4} + \etakb \,,
\ee 
where we have set $L=R$ in the definitions of \Eqref{etas}. These interfaces are, however, always unstable. As $\nabla_{\bm N}(1/R) = 0$, \Eqref{stability} reduces to \Eqref{negative KG} also for non-zero $\etak$ and $\etakb$. This condition is clearly never satisfied on the sphere, thus, for non-conserved area fractions, spherical vesicles cannot support interfaces. In practice, this implies that a multicomponent scaffolded lipid vesicle allowed to exchange lipids with the environment, will eventually expel the stiffer phase (i.e. the phase having the largest elastic moduli).

For conserved area fractions, on the other hand, one can demonstrate that CGC lines become stable, as the second variation of the free energy is
\be 
\delta^{(2)} F
=
\frac{2\sigma }{R |\sin \psi|} 
\sum_{n>0}
|\epsilon_n|^2 
\left(
n^2 
- 1
\right) 
\,,
\ee
with $\epsilon_{n}$ the amplitudes of the Fourier components of a small displacement along the tangent-normal direction, is always non-negative (see Appendix \ref{app:second variation}). As in any conserved order parameter system, Lagrange multipliers remove the zero mode instabilities.

Although CGC lines are always stable on the sphere, configurations featuring multiple domains are inevitably local minima of the free energy, whereas the configuration consisting of a single hard and a single soft domain is the global minimum. To prove this statement, we calculate the difference in free energy between a configuration comprising $N$ circular identical domains, each of fractional area $\varphi/N$, and single circular interface. This yields:
\be
\frac{F_{N} - F_{1}}{4 \pi \sigma R }
=
\sqrt{\varphi  (N- \varphi)} - \sqrt{\varphi (1- \varphi) }\;.
\ee
which does not depend on the bending moduli and is positive for any $\varphi$ and $N>1$.  For this reason, as in flat geometries, a single interface will be always preferred on a spherical substrate. These considerations evidently do not apply to GUVs, where multiple circular domains are routinely observed, see e.g. \cite{veatch2005seeing}. This can be ascribed to the budding of phase domains \cite{Kuzmin2005}, although other stabilization mechanism have also been proposed \cite{Frolov2006}.

\subsection{Axisymmetric surfaces}

\label{sec:axisymmetric}

The full rotational symmetry of the sphere results into a mere renormalization of the chemical potential, but does not provide the prerequisite for a geometry-induced localization of lipid domains (i.e. geometric pinning).  In order to appreciate the effect of the underlying geometry, one has to consider surfaces with non-uniform curvature. 

The simplest way to achieve this is to consider surfaces which are invariant under the isometries of Euclidean space, namely rotations and translation. In this section we discuss the case of surfaces equipped with an axis of rotational symmetry (i.e. axisymmetric surfaces or surfaces or revolution) and in Sec. \ref{sec:developable} we extend our analysis to developable surfaces, which represent a larger class that includes translationally invariant surfaces. Due to their simplicity, axisymmetric surfaces have played a special role in the membrane physics literature, starting from the early work of Helfrich and Deuling \cite{deuling1976curvature} and Jenkins \cite{jenkins1977static}. In the context of phase-separated domains on membranes, they were the only class of surfaces studied in Ref. \cite{Julicher1996}, as well as the only class used to compare theory and experiments in \cite{Baumgart2005}.

\begin{figure}
\centering
\includegraphics[width=\columnwidth]{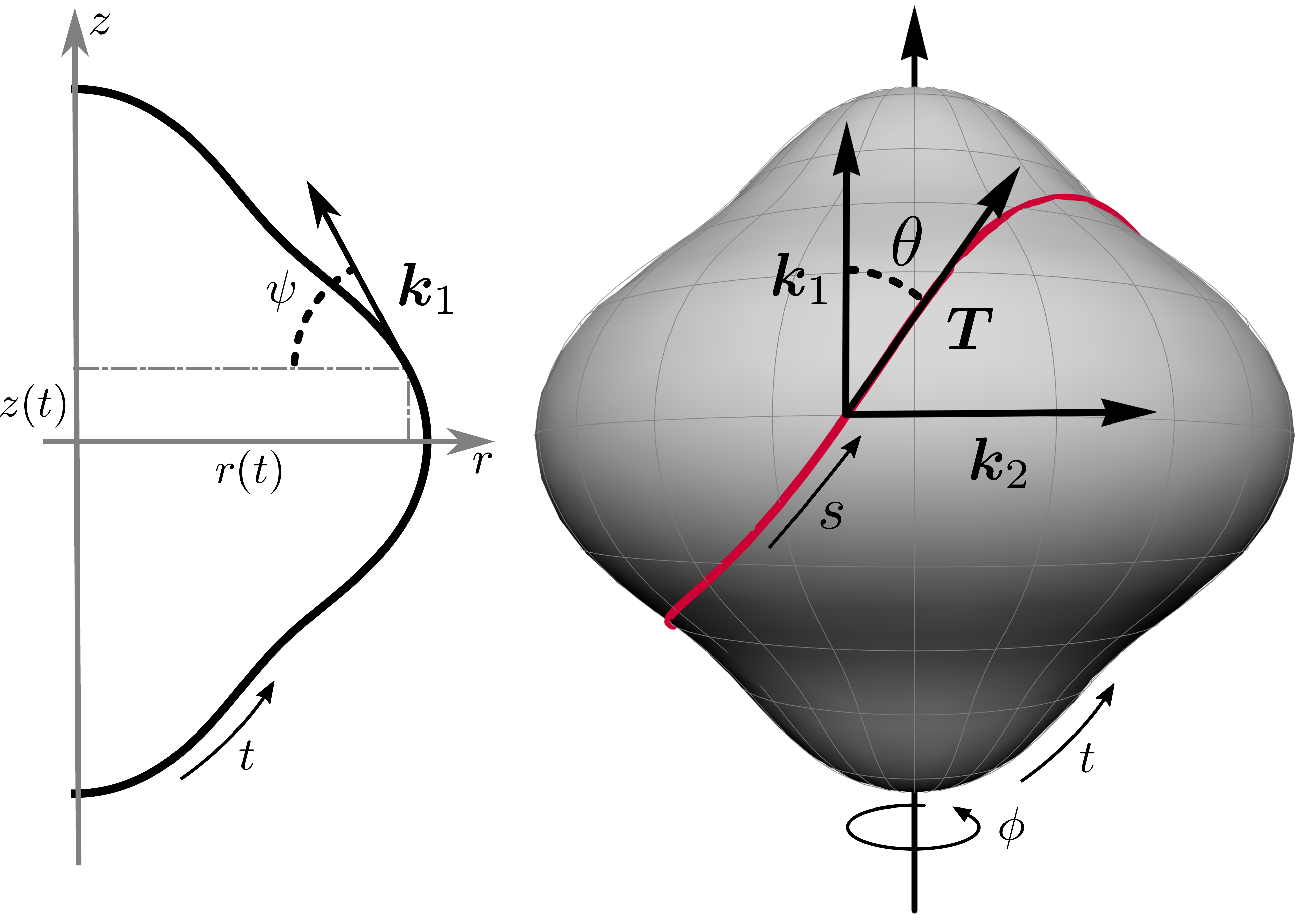}
\caption{Parametrization of an axisymmetric surface. On the left, we show the radial profile $\{r(t),z(t)\}$ as function of $t$, the arc-length parameter of the profile. The full surface is obtained by rotations along the $z$ axis parametrized by the angle $\phi$, as in \Eqref{axisymmetric parametrization}. We define $\psi=\psi(t)$ to be the angle between $\bm{k}_1$, the meridian principal direction, and the horizontal plane. On the right, we show how the curve $\Gamma$ on $\Sigma$ is parametrized in its own arc-length, $s$: its unit tangent vector $\bm{T}$ makes an angle $\theta=\theta(s)$ with $\bm{k}_1$. Notice that the orientation of the principal directions is not fixed a priori: we choose it to match the one of $\{\bm{T},\bm{N}\}$.} 
\label{fig:spaccato}
\end{figure}

\begin{figure*}
\includegraphics[width=\linewidth]{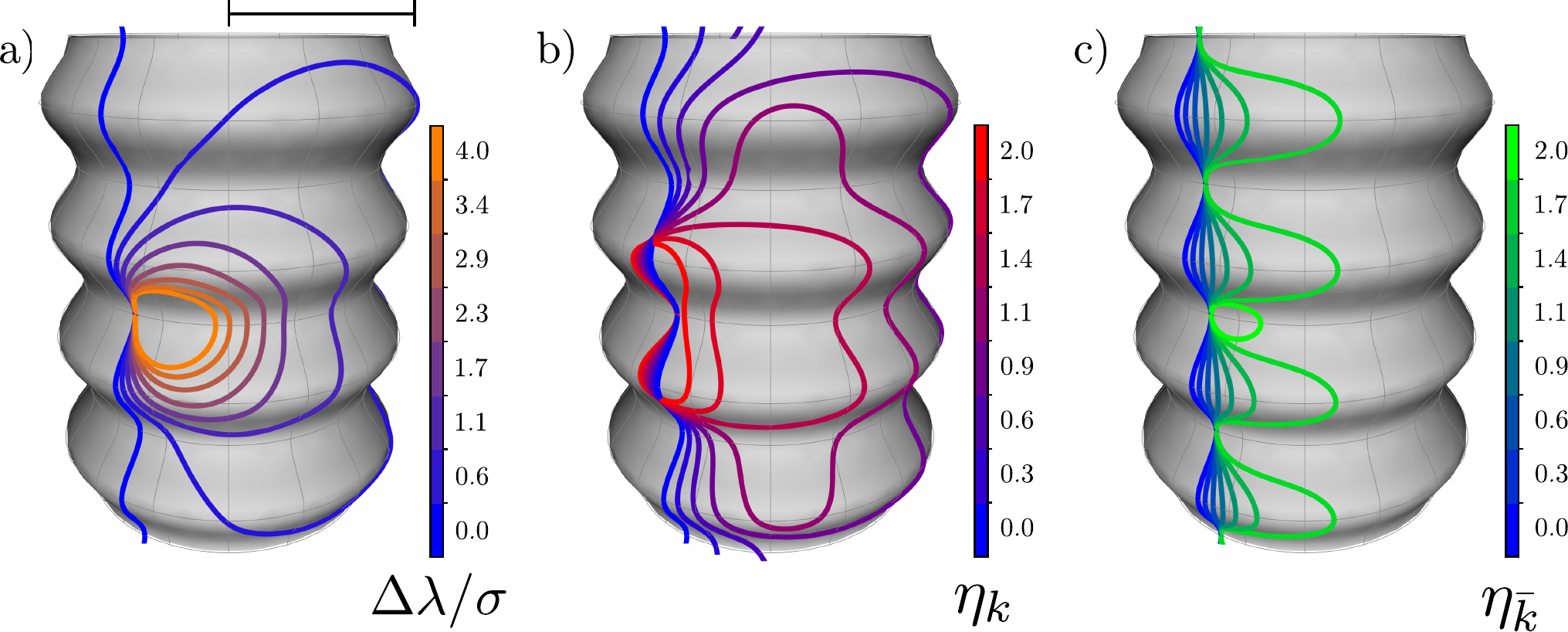}
\caption{
Solutions of \Eqref{axisymmetriceq} on a corrugated cylinder. \textbf{a)} Interfaces with varying $\Delta \lambda$: they are CGC lines, and look like circles. \textbf{b)} Curves with varying $\eta$: interfaces are closed but the shape is non-circular. \textbf{c)} Varying the coupling with the Gaussian curvature: closed curves are possible only for high values of $\etakb$ and do not encompass multiple corrugations. In all three panels the pure 	blue vertical line has zero coupling and thus correspond to a vertical geodesic. The scale bar on the top left shows the value chosen for $L$ in \Eqref{etas}.}
\label{fig:tube}
\end{figure*}

Rotationally invariant surfaces are completely characterized by their radial profile. 
Choosing $\bm{\hat{z}}$ as symmetry axis, one can parametrize arbitrary axisymmetric surfaces as:
\be 
\bm{r}(t,\phi) =  \left\{ r(t) \cos \phi , r(t) \sin \phi , z(t) \right\} \,,
\label{axisymmetric parametrization}
\ee
where $t$ is the arc-length parameter of the cross-section and $\phi \in [0,2\pi]$ is the usual polar angle on the $xy-$plane (see \fref{fig:spaccato}). The mean and Gaussian curvatures are then given by:
\be
H = -\frac{1}{2}\frac{\D\psi}{\D t} -\frac{ \sin \psi }{2 r }\;, \qquad 
K = \frac{\sin \psi }{r}\,\frac{\D\psi}{\D t}\;,
\label{eq:axisymmetric_curvatures}
\ee
where $\psi=\psi(t) = \arctan (\D z/\D t)/(\D r/\D t)$, is the angle between the meridian direction $\bm{k}_1$ and the constant $z-$plane, see \fref{fig:spaccato}. When evaluated along $\Gamma$, both curvatures are functions of the arc-length coordinate $s$. The principal directions coincide with parallels and meridians. The latter, in particular, are also geodesic (they indeed are the shortest path between points with the same angular coordinate $\phi$), hence have vanishing geodesic curvature. On the other hand, the parallels have in general non-zero geodesic curvature
\be
\kappa_g(\bm{k}_2) = \frac{\cos \psi}{r}  \,.
\ee
A sphere of radius $R$ would have $r=R \sin \psi$, and the above expression recovers \Eqref{CGC sphere}.

More in general, a curve $\Gamma$ on an axisymmetric surface is parametrized by a pair of functions $\{t(s),\phi(s)\}$. Its geodesic curvature can be expressed in the form
\be 
\kappa_g 
= \dot{\theta} + \frac{\sin \theta \cos \psi}{r}
= \frac{1}{r}\frac{\D}{\D t}\left(r\sin\theta\right)\;,
\label{eq:axisymmetric_kg}
\ee
where $\theta=\theta(s)$ is the angle between the tangent vector of $\Gamma$ and the local meridian, so that $\bm{T}=\dot{\bm{r}}=\cos \theta\,\bm{k}_{1}+\sin \theta\,\bm{k}_{2}$. \Eqref{eq:axisymmetric_kg} implies the so called {\em Clairaut relation}, according to which geodesics with $\theta \neq \pi/2$ on axisymmetric surfaces satisfy:
\be
r \sin \theta = \mathrm{const} \,.
\label{eq:clairaut}
\ee
In particular meridians, whose tangent vector is parallel to $\bm{k}_1$, have $\theta=0$ and are thus geodesics. Using Eqs. \eqref{eq:axisymmetric_curvatures} and \eqref{eq:axisymmetric_kg}, the interface \Eqref{interfaceq} can be cast in the form:
\begin{multline}
\frac{1}{r}
\frac{\D}{\D t}\left[ r \sin \theta + \left( \frac{\Delta \bar{k}}{\sigma}+ \frac{\Delta k}{2\sigma}\right) \cos \psi \right]\\[5pt]
=
\frac{\Delta k}{4\sigma}
\left[
\left(\frac{\D\psi}{\D t}\right)^{2}
+
\frac{\sin^2 \psi}{r^2} 
\right]
+
\frac{\Delta\lambda}{\sigma}
 \,.
\label{axisymmetriceq}
\end{multline}

In principle, this interface equation is integrable, since it can always be put in the generic form
\begin{equation}
\frac{1}{r}\frac{\D}{\D t} \left[ r \sin \theta + f(t) \right]=0\;,
\end{equation}
with a properly chosen $f(t)$. The quantity between brackets is a constant of motion whose conservation is a direct consequence of the rotational invariance of the surface. For generic couplings $\etak$, $\etakb$, $\Delta \lambda$ finding such $f(t)$ amounts to finding the $t-$primitive function of the right-hand side of \Eqref{axisymmetriceq}, which is not always possible analytically and thus is not particularly useful. However, if $\Delta\lambda=0$ and there is no coupling with the mean curvature (i.e. $\Delta k = 0$), we find the relation
\be 
r \sin \theta + \frac{\Delta \bar{k}}{\sigma} \cos \psi = \mathrm{const.}
\label{axisymmetricminimal}
\ee
which is true for any $r(t)$ and could be viewed as a generalization of the Clairaut relation \Eqref{eq:clairaut}. The value of the constant is fixed by boundary conditions. In fact, if $\Sigma$ is a catenoid, which is the only axisymmetric surface such that $H=0$ everywhere, then \Eqref{axisymmetricminimal} is the most general interface equation for a non-fixed area fraction.

\begin{figure*}[t!]
  \includegraphics[width=\linewidth]{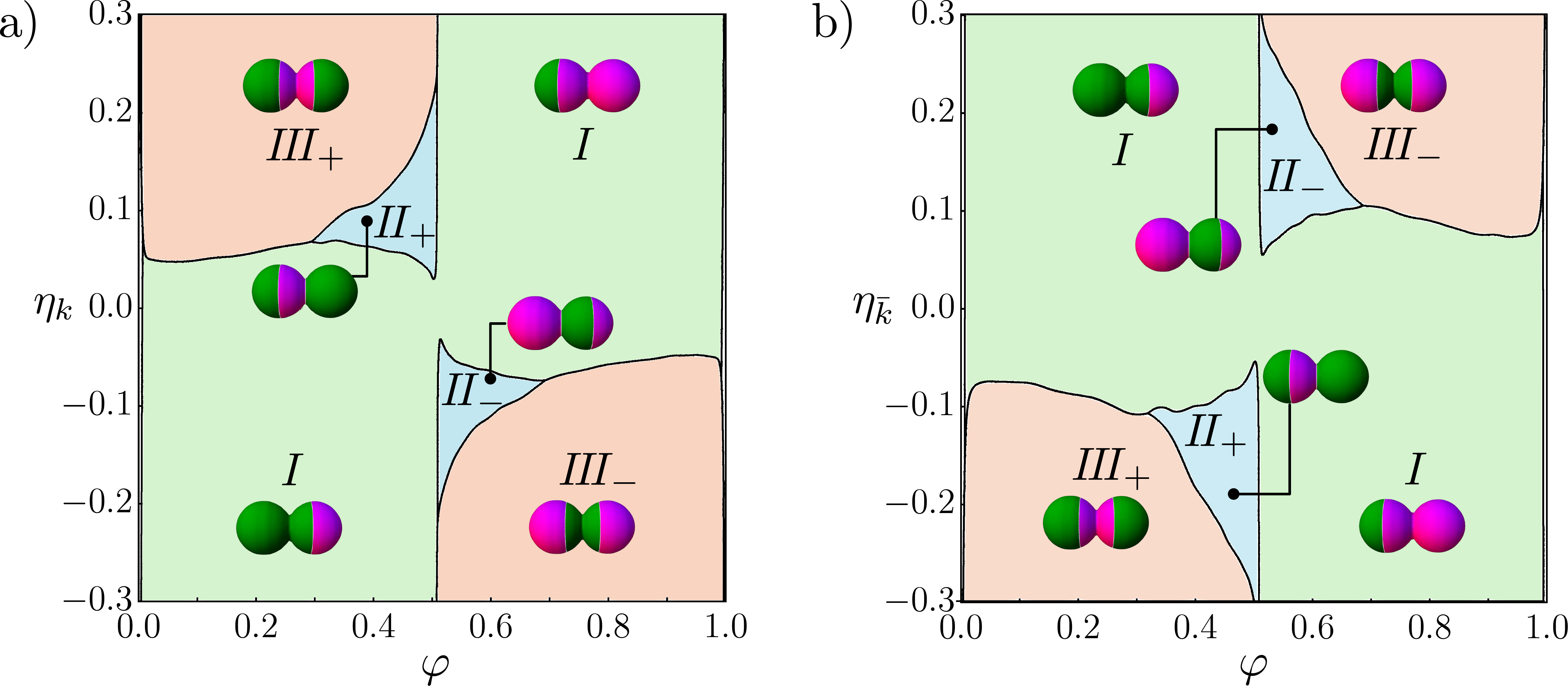}
  \caption{Phase diagrams of minimal configurations for the dumbbell-shaped particle of \fref{fig:curves on db}, for varying $\varphi$. \textbf{a)} The effect of $\etak$ while keeping $\etakb=0$. \textbf{b)} The effect of $\etakb$ while keeping $\etak=0$. In both panels, different colors correspond to different minimal energy configurations: type $I$ (light green) consist of two domains and one interface; types $I\!I_\pm$ (light blue) and type $I\!I\!I_\pm$ (light red) have two interfaces and three domains. In the insets, hard and soft phases are respectively depicted in green and magenta. All interfaces are CGC parallels. We set $L=\sqrt{A_\Sigma}$.}
\label{fig:db phase diagram}
\end{figure*}

\fref{fig:tube} shows solutions of \Eqref{axisymmetriceq} for a corrugated cylinder, i.e. a cylinder with a periodic wave-like perturbation along the axial direction. Compared to \fref{fig:local surfs}, this geometry better highlights the effect of $\Delta \lambda$, $\etak$ and $\etakb$ on the structure of the interface. The fact that both $H$ and $K$ are non-constant along the axial direction, in particular, leads to highly non-trivial interface geometries. For simplicity, here we consider only interfaces whose tangent vector is parallel to the $z-$axis at least at one point. For $\Delta \lambda=\etak=\etakb=0$, the interfaces are then vertical geodesics (pure blue vertical curves in \fref{fig:tube}). For $\Delta \lambda \neq 0$, but $\etak=\etakb=0$, these are CGC lines (\fref{fig:tube}.a), whose shape clearly resembles that of a circle. For $\etak \ne 0$ and $\Delta\lambda=\etakb=0$, on the other hand, the interfaces become more elongated and extend to multiple ``valleys'' (\fref{fig:tube}.b). Finally, for $\etakb \ne 0$ and $\Delta\lambda=\etak=0$ (\fref{fig:tube}.c), the solutions of \Eqref{axisymmetriceq} are either deformations of vertical geodesics or small circles sitting in a single valley.

In general, for any given substrate geometry, there exists a plethora of possible solutions of \Eqref{interfaceq}. To gain insight on the physical mechanisms underlying geometric pinning in axisymmetric surfaces and make a connection to the experiments \cite{Rinaldin2018}, here we make the further assumption that, like the substrate, the interface is also rotationally invariant. Then, for conserved area fractions, {\em every} parallel is a solution of \Eqref{interfaceq} for a specific $\varphi$ value, regardless of the values of $\etak$ and $\etakb$. The problem thus reduces to finding a configuration of domains that minimizes the free energy. 

Intuitively, for small $\etak$ and $\etakb$ the force balance is dominated by line tension. Thus the system partitions in two domains separated by a single interface, whose position is trivially determined by the area fraction. Upon increasing  $\etak$ and $\etakb$, on the other hand, configurations featuring multiple domains might become energetically favoured. We stress that the number of domains alone is not necessarily a good indicator of the strength of geometric pinning, as complex substrate geometries (such as the corrugated cylinder of \fref{fig:tube}) can allow for stable equilibria with multiple domains even when $\etak=\etakb=0$. In this respect, curved and flat substrates are dramatically different from each other, as on flat substrate interfaces are always circular (or straight as a limiting case). 

As a concrete example, in \fref{fig:db phase diagram} we show the phase diagram of a dumbbell-shaped binary vesicle (the shape of $\Sigma$ is precisely the one of \fref{fig:curves on db}), such as that we have experimentally studied in Ref. \cite{Rinaldin2018}. In the left panel, we set $\etakb=0$ while varying the area fraction $\varphi$ and $\etak$, while in the right panel we vary $\etakb$ and keep $\etak=0$. 

We then proceed to compare the total energy of different types of configurations, here labelled $I$, $I\!I_\pm$ and $I\!I\!I_\pm$. In the insets, the ``$+$'' domains are coloured in green and the ``$-$'' domains in magenta. Type $I$ is the configuration consisting of only two domains, separated by a single interface. Types $I\!I_{\pm}$ and $I\!I\!I_{\pm}$ consist of three domains and two interfaces.  Configurations $I\!I_{\pm}$ have always one interface lying along the dumbbell	 neck (where the interface is shortest), while the second interface varies according to the value of $\varphi$. Configurations $I\!I\!I_{\pm}$ have instead two symmetrical interfaces at the same distance from the neck region.

As expected, for $\eta_{k}=0$ the only stable configuration consists of only two domains separated by a single interface (type $I$). However for $\etak > 0$ we see that three-domain configurations can become favourable when $\varphi < 0.5$. Similarly, for $\etak < 0$ we find that three domains become favourable for $\varphi > 0.5$. This symmetry of the phase diagram is a direct consequence of the fact that the free energy, \Eqref{freeenergy}, is invariant under the transformation $\etak \to -\etak$ and $\varphi  \to 1-\varphi$. The right panel shows that the situation for non-zero $\etakb$ is reversed: in order for the configuration $I\!I\!I_+$ to become energetically favourable, we need to have $\etakb<0$. Interestingly, the type $I\!I\!I_-$ has been observed in experiments \cite{Rinaldin2018} and thus points out to the fact that for the real membranes $\etak$ and $\etakb$ likely have opposite signs.

\subsection{Minimal surfaces}

\label{sec:minimal}

Minimal surfaces are surfaces with zero mean curvature. These surfaces locally minimize both the area and the bending energy and are, therefore, commonly found in nature in a variety of systems, including self-assembled lipid structures in water or water/oil mixtures \cite{Andersson1988}. 

\begin{figure*}[t]
\includegraphics[width=\linewidth]{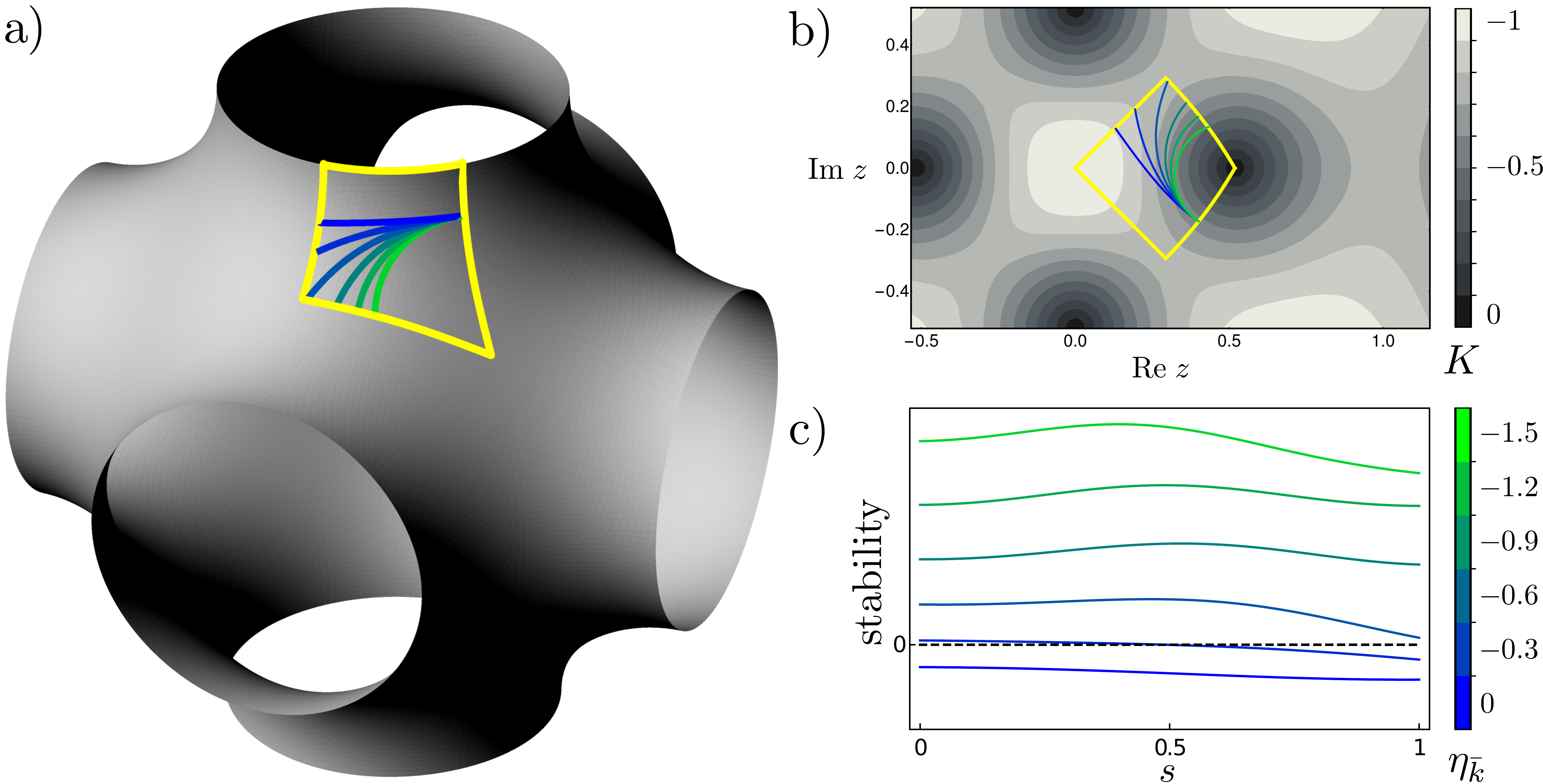}
\caption{ 
Interfaces on the Schwarz P surface. \textbf{a)} We highlight in yellow the boundary of the fundamental patch. By properly gluing 48 copies of this patch one can construct a full unit cell, here drawn by aid of Surface Evolver \cite{Emmer1992}.
\textbf{b)} The interface \Eqref{eq:WE interface} is actually solved for a curve in $\mathbb{C}$, and solutions are shown as blue/green curves, with blue corresponding to geodesics. The yellow lines highlight the region $\mathcal{D}$ (defined in \Eqref{WE D}) which is mapped onto the fundamental patch in a). The gray contour shades in the background show the value of the Gaussian curvature $K$ (see  \Eqref{WE K}) as a function of $z$. The value of the curvature is rescaled so that the range shown is $[0,-1]$. Note that zeroes of $K$ correspond to poles of $f(z)$. All solutions have identical initial conditions: they start off from them same point on the boundary of the patch and have initial tangent vector $T^i(0)$ such that $\bm{T}(0)$ is pointing horizontally in the embedded surface, as is clear from a). Note that in principle all complex curves extend without problem outside $\mathcal{D}$, but cannot be mapped onto the surface. 
\textbf{c)} Evaluation of \Eqref{eq:WE stability} along the solutions displayed in a) and b). Each curve parameter $s$ is rescaled so that it spans the interval $[0,1]$. The vertical axis uses arbitrary units, since only the sign of the stability factor contains relevant information. By increasing the modulus of $\etakb$ we see that interfaces become progressively more unstable; namely only $\etakb=0$ (geodesic) and the $\etakb=-.3$ seem to allow locally stable interfaces. 
}
\label{fig:figure schwarz}
\end{figure*}

As $H=0$ everywhere, the free energy of a multicomponent vesicle, \Eqref{freeenergy}, can be expressed as a contour integral along the interface only, by virtue of the Gauss-Bonnet theorem. This yields:
\be 
F = \int_\Gamma \D s \left(\sigma  -\Delta \bar{k} \, \kappa_g \right) + 2 \pi \Delta \bar{k} \chi_{+}\,,
\ee
where $\chi_{+}$ is the total Euler characteristic of the $\Sigma_+$ domains. The Gaussian curvature is always non-positive, and every non-planar point of the surface is saddle-like. Since any closed surface of finite area is required to have some regions with $K>0$ (see e.g. \cite{DoCarmo1976}), there cannot be compact minimal surfaces without boundaries. Nevertheless, several systems adopt a minimal configuration which extends for a finite size, eventually stopping at some boundary regions or repeating periodically. In the following we assume that we can ignore any sort of boundary effect and will focus on portions of the surface where the minimality condition holds.

The Gaussian curvature $K$ can be evaluated on the curve using only quantities relative to the Darboux frame, so that \Eqref{interfaceq} becomes
\be 
\kappa_g + \etakb \left(\kappa_n^2 + \tau_g^2 \right) = \frac{\Delta \lambda}{\sigma} \,, 
\label{eq:minimal interface}
\ee
with $\kappa_n$ and $\tau_g$ the normal curvature and the geodesic torsion of $\Gamma$, defined in Appendix \ref{app:curves}. The length scale $L$ used in the definition of $\etakb$ here corresponds to the overall size of the surface, or, in the case of periodic surfaces, to the surface wavelength. If $\varphi$ is not conserved, the right-hand side of \Eqref{eq:minimal interface} vanishes and we have that $\kappa_g=-\etakb(\kappa_{n}^{2}+\tau_{g}^{2})$. Thus, the concavity of the interface is solely determined by the sign of $\etakb$. Since $\etakb$ is usually negative \cite{Baumgart2005,Rinaldin2018}, this means that the interface will form convex domains of the \textit{soft} phase.
However, the non-trivial topology and geometry of minimal surfaces might counter this intuition.

In any case, even if the formation of closed domains is possible, the interface needs to be stable, which for minimal surfaces amounts to satisfy the condition
\be
K \left(1+ \etakb^2 K \right) + \etakb \nabla_{\bm{N}} K \leq 0\;,
\label{eq:minimal stability}
\ee
which depends only on the value of the Gaussian curvature and its normal variation at any given point of $\Gamma$.
For small and negative $\etakb$, this inequality implies that soft domains are likely to be stable in regions of high $|K|$, and conversely, hard domains might be more stable in regions were $|K|$ is small.

Although expressed in a compact form, both \Eqref{eq:minimal interface} and inequality \Eqref{eq:minimal stability} do not allow to easily extract further physical information and are not well suited for numerical solutions. To overcome this, we use of the well-established Weierstrass-Enneper (WE) representation  (see e.g. Ref. \cite{Colding2011}) to parametrize generic minimal surfaces as harmonic maps (see Appendix \ref{app:weierstrass} for details). This representation has several advantages, including the fact that it naturally selects \textit{isothermal} coordinates, i.e. coordinates in which the metric over $\Sigma$ is conformally flat.

If the surface is described as an explicit embedding $\bm{r}(u,v)$, we can combine the two parameters $\{u,v\}$ into a single complex variable $z=u+i v$. Then, a curve on the surface, parametrized as $\bm{r}(s)=\{u(s),v(s)\}$, can be seen as a complex curve $z(s) \in \mathbb{C}$ mapped onto $\mathbb{R}^3$. Consequently, the interface \Eqref{eq:minimal interface} can be rewritten as a first order differential equation for a curve over the complex plane: 
\be
\dot{\alpha}
+ \frac{2}{\Omega} \im e^{i \alpha} \partial_z \log \Omega 
+\etakb \frac{4 |f g_z|^2}{\Omega^4} = \frac{\Delta\lambda}{\sigma} \,,
\label{eq:WE interface}
\ee
where $\alpha=\alpha(s)$ is such that $\bm{T}= \cos\alpha \, \bm{t}_u + \sin \alpha \, \bm{t}_v$, with $\{\bm{t}_u,\bm{t}_v\}$ the tangent vectors in the $u$- and $v$-directions. The quantity $\Omega=|f|(1+|g|^2)$ is the conformal factor appearing in the induced metric, $f=f(z)$, $g=g(z)$ are the two complex WE functions, and $g_z=\partial_z g$. Similarly, the stability condition for non-conserved $\varphi$, \Eqref{eq:minimal stability}, becomes equivalent to
\be 
1 + \frac{2 \etakb}{\Omega} \im e^{i \alpha} \partial_z \log \frac{f g_z}{\Omega^4} -4 \etakb^2 \frac{|f g_z|^2}{\Omega^4} \geq 0 \,.
\label{eq:WE stability}
\ee
The overall phase of $f(z)$ is usually treated as an independent parameter, called the \textit{Bonnet angle} $\theta_B$. Neither the interface equation nor the stability condition depend on it. In fact, different values of $\theta_B$ correspond to different immersions of the same intrinsic geometry: these immersions are locally isometric to each other and define a family of surfaces, called the Bonnet family. Clearly, both \Eqref{eq:WE interface} and \Eqref{eq:WE stability} hold equally for all members of the same Bonnet family.

For instance, the catenoid and the helicoid belong to the same family, as they can be continuously mapped into each other, and both have WE functions $f(z)=e^z/2$ and $g(z)=e^{-z}$. By plugging these values into \Eqref{eq:WE interface} one can obtain a very compact expression for the interface equation which can be easily solved numerically, and then one can use \Eqref{eq:WE stability} to check stability of solutions.

We choose to focus instead on another class of surfaces which is of much greater physical importance. These are the \textit{triply periodic minimal surfaces} (TPMS), a type of periodic structures which extend and repeat infinitely in all directions and divide the full space into two distinct, non-intersecting and mutually interwoven labyrinth systems. Several examples of such surfaces are known, three of which have been extensively observed and studied in self assembled lipid structures over the past decades \cite{Hyde1984,Seddon2000,Kaasgaard2006,Demurtas2015}. Such peculiar surfaces occur also in biological systems, e.g in mammalian lung tissue \cite{Larsson2014} or inside mitochondria \cite{Deng2009}.

\begin{figure}[t]
  \includegraphics[width=\columnwidth]{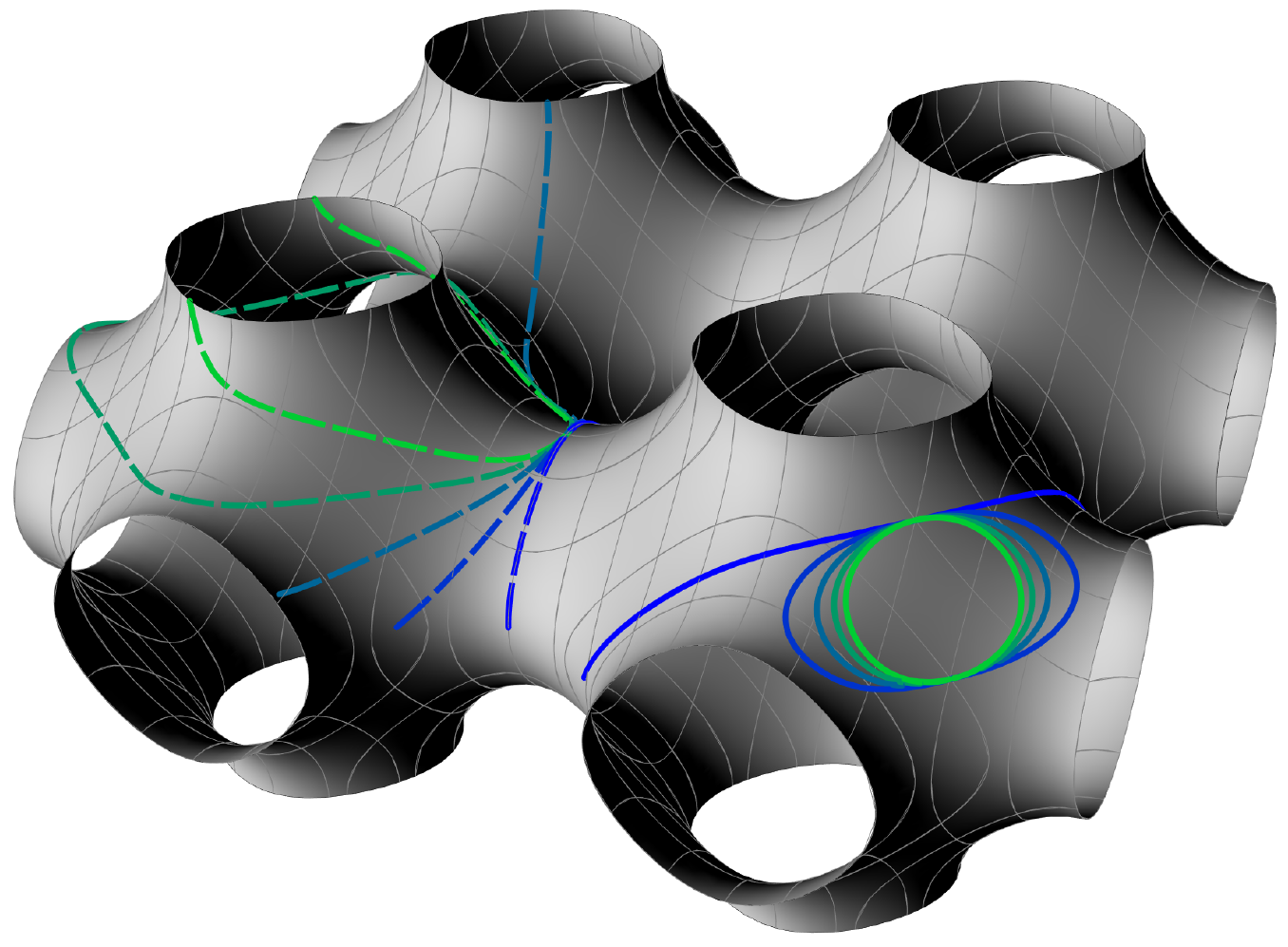}
  \caption{The layers $z_{+,0}$ and $z_{-,1}$ (see \Eqref{nodal layer}) of the  nodal approximation of the P surface form together a planar square lattice of unit cells, four of which are displayed. In the upper layer we show some solutions of \Eqref{eq:minimal interface}. The color coding of the continuous lines is the same as that of \fref{fig:figure schwarz}, while the dashed lines have $\etakb$ rescaled by a factor $1/10$. We evaluated \Eqref{eq:minimal stability} for the solutions, and found that, for the  continuous curves, only the geodesic and the $\etakb=-.3$ are stable, similarly to what was found in \fref{fig:figure schwarz} for the exact solution. We postulate that for the latter value of $\etakb$ it is possible to have the formation of stable, finite size soft domains on the P surface. The stability of the dashed lines is much more $s$-dependent. The first three curves seems to be stable: they correspond to non-contractible closed interfaces encompassing several unit cells.
  }
\label{fig:figure nodal}
\end{figure}

These three minimal surfaces are known as \textit{gyroid} and Schwarz P and D surfaces, and are extensively discussed in Appendix \ref{app:schwarz P}. They all belong to the same Bonnet family, thus we will restrict the following discussion to the case of the P surface, even if every result we obtain can be generally applied to any of the three.

The WE functions for the P surface are $f(z)=(1-14 z^4 +z^8)^{-1/2}$ and $g(z)=z$, defined on a region of $\mathbb{C}$ known as the \textit{fundamental patch} (the region highlighted in yellow in \fref{fig:figure schwarz}). The full surface is constructed by gluing together different properly oriented patches: it takes 48 of them to form the unit cell of \fref{fig:figure schwarz}.a. The unit cell then repeats periodically in all three directions to form a cubic lattice. It is possible to give an analytic expression for the embedding of the P surface in terms of incomplete elliptic integrals \cite{Gandy2000}.

Within a single patch, we are able to solve the WE interface \Eqref{eq:WE interface} and evaluate the stability condition \Eqref{eq:WE stability} on the solutions. Some exemplificative results are shown in \fref{fig:figure schwarz}. For $\etakb=0$ we find that geodesics are always stable, in accordance with the general discussion on geodesic interfaces of Section \ref{sec:lines on surfaces}. We discover that, upon increasing the modulus of $\etakb$, interfaces quickly become more unstable. In fact, regardless of the direction of the interface on the patch, for $|\etakb| \gtrsim 2$ we have never been able to observe a stable interface. 

Conversely, for milder curvature couplings we find that stable solutions exist. Predicting whether these correspond to closed, simply-connected lines, and thus can serve as viable interfaces for finite domains, is complicated by the fact that these curves naturally encompass several patches, while the WE representation is well-defined only on a single patch. Since the gluing conditions for a curve travelling throughout the surface are non-trivial, we opted for an alternative method: we used the so called \textit{nodal approximation} \cite{VonSchnering1991} of the surface, described in Appendix \ref{app:schwarz P}. For instance, this approximation was successfully used in Ref. \cite{Demurtas2015} to mathematically model observations done with electron microscopy.

The nodal surface has the same space group of the P surface and has the crucial advantage that it can be easily expressed as a (stack of) vertical graphs defined over a whole lattice plane. Equivalently, it admits a very easy representation in terms of functions of the form $z(x,y)$, where $x$ and $y$ are two lattice axes. This surface is not exactly minimal, but $H^{2} \ll |K|$ everywhere. Therefore, we cannot use the WE construction to solve the interface equation, but have to rely on the general \Eqref{eq:minimal interface}. Although more tedious, we managed to find numerical solutions, as shown in \fref{fig:figure nodal}.

We find that, for the same $\etakb$ values shown in \fref{fig:figure schwarz}, the system does admit closed interfaces and, using \Eqref{eq:minimal stability}, we find that some of these are stable, provided $|\etakb|$ is not too big. In particular, the outermost closed continuous blue curve in \fref{fig:figure nodal} is stable, whereas the others (in green) are not. Even milder values of the coupling can lead to topologically non-trivial interfaces encompassing several unit cells, as shown by the dashed lines in \fref{fig:figure nodal}. However, assessing the stability of these curves is a more delicate procedure and likely the nodal approximation cannot be trusted entirely. 

Phase separation on the P surface was previously studied in Ref. \cite{paillusson2016phase}, using a discrete Ising model coupled to the Gaussian curvature $K$. The key difference with the present results is that the analysis reported in Ref. \cite{paillusson2016phase} focuses on a single unit cell with conserved $\varphi$. Whereas the conservation of area fraction is likely a global property of cubic systems, this might not be necessarily the case at the scale of a single unit cell.

\subsection{Developable surfaces}

\begin{figure}[t]
  \includegraphics[width=\columnwidth]{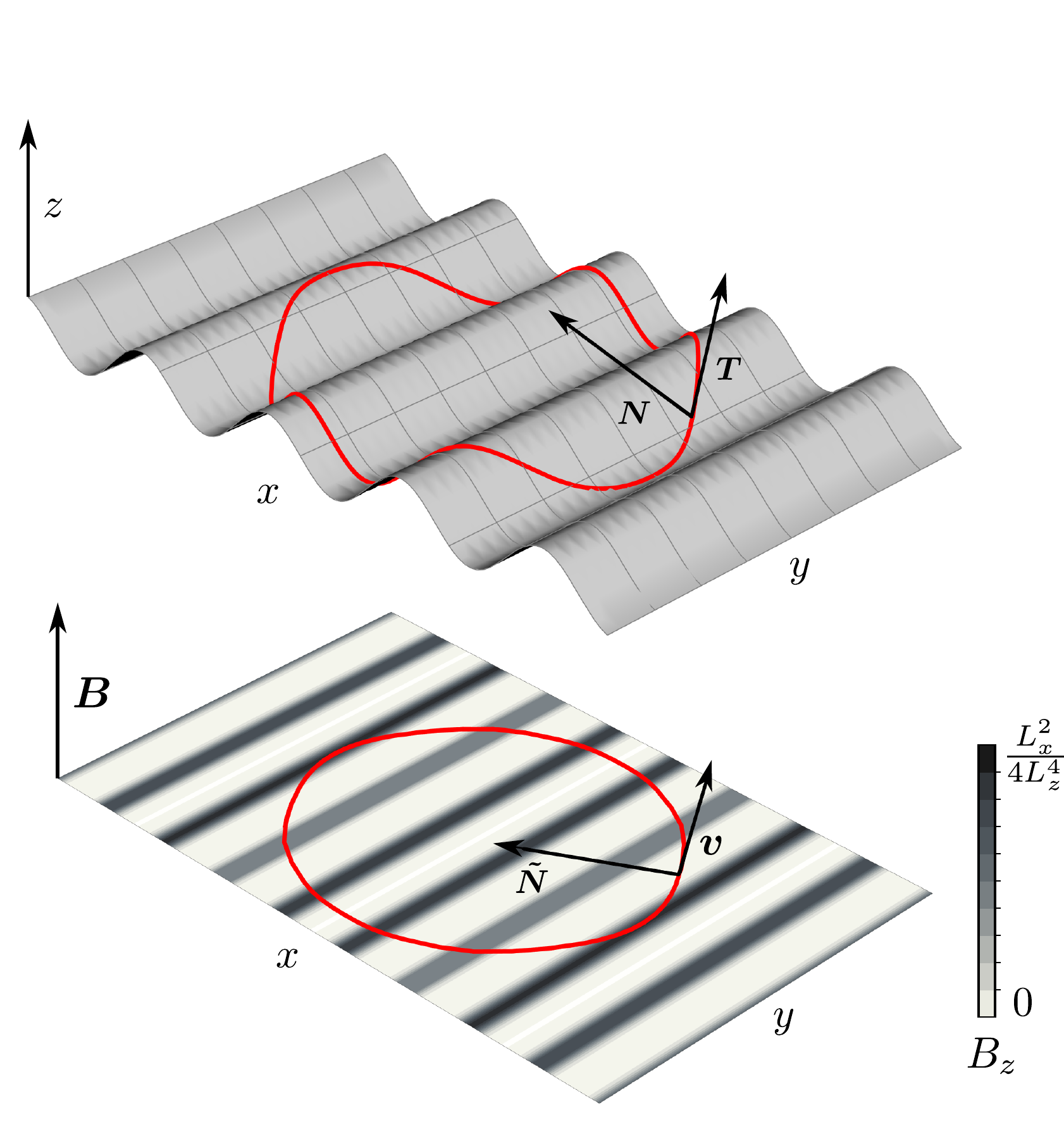}
  \caption{
The analogy between mean curvature and magnetic field. The top panel shows a developable surface with translational invariance and sinusoidal profile height $z(x)=L_z \sin x/L_x$. The red curve is a generic interface with $\Delta \lambda L_x=\sigma/2$ and $\etak=1/10$, obtained with initial conditions $x_0=0$ and $\theta_0=\pi/2$. This closed curve is analogue to the \textit{planar} trajectory of a charged particle in a $x$-dependent axial magnetic field $\bm{B}=B_z \bm{\hat{z}}$, which oscillates between the values $\Delta \lambda/\sigma$ and $\Delta \lambda/\sigma+\etak L_x^2/4L_z^4$ with spatial periodicity of $\pi L_x$. The tangent $\bm{T}$ is mapped to the planar velocity $\bm{v}$, and the normal $\bm{N}$ is mapped to in-plane normal $\bm{\tilde{N}}$.}
\label{fig:figure developable 1}
\end{figure}

\label{sec:developable}

Developable surfaces are those having everywhere vanishing Gaussian curvature. By virtue of Gauss' {\em theorema egregium}, they can be isometrically mapped onto a plane (see Appendix \ref{app:surfaces}). Cylinders, cones, developable ribbons \cite{Hinz2015,wunderlich1962abwickelbares,Giomi2010} and surfaces which are invariant under a rigid translation, as the corrugated substrates experimentally studied in Ref. \cite{parthasarathy2006curvature} and described in Ref. \cite{Rozycki2008}, are all common examples of developable surfaces. Curves embedded on developable surfaces are simpler to describe than in the general case: with trivial intrinsic geometry, lines of curvature are also geodesics and the geodesic curvature of an arbitrary curve is simply $\kappa_g= \dot{\theta}$, as in flat space. Thus, geodesics on such surfaces always make a constant angle with principal directions.

\begin{figure*}[t]
  \includegraphics[width=\linewidth]{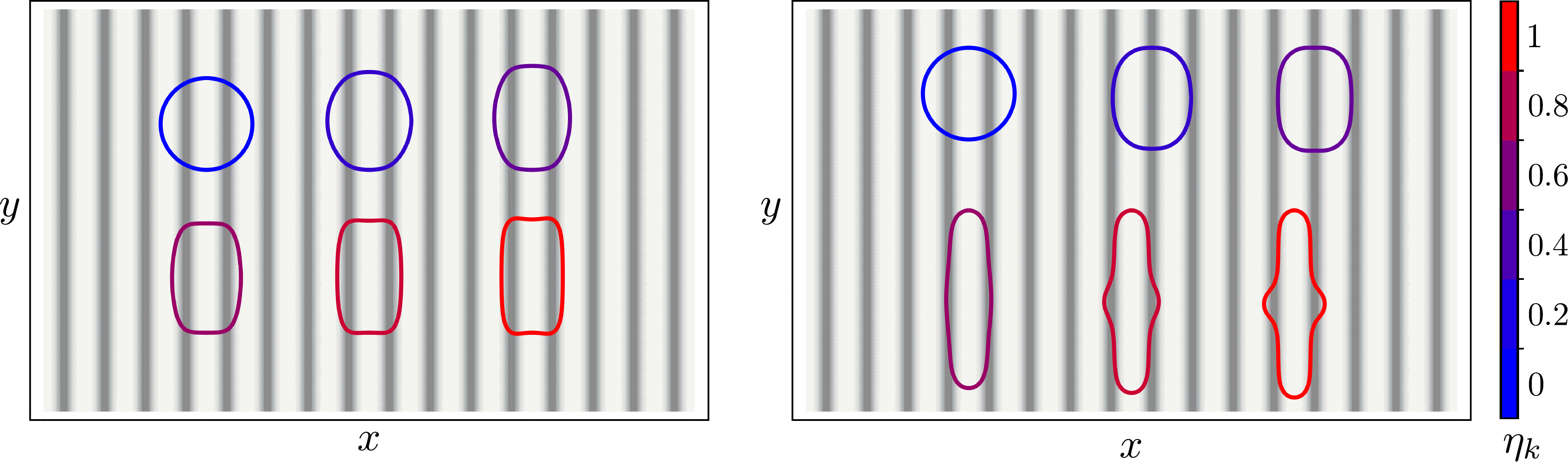}
  \caption{
  Using the same sinusoidal surface geometry of \fref{fig:figure developable 1}, we display closed interfaces which minimize \Eqref{dev flux} for varying $\etak$. Each interface is chosen such that the area of the enclosed domain is equal to $L_x^2$ (and we set $L=L_x$ in \Eqref{etas}). The left panel shows $\Sigma_-$ domains, i.e. soft domains surrounded by a LO background. The right panel shows $\Sigma_+$ domains, i.e. hard domains surrounded by a LD background. Gray tones in the background follow the same color scheme of \fref{fig:figure developable 1}, indicating the magnitude of $\bm{B}$. While for each $\etak$ there are multiple solutions which have the same area, only the ones minimizing \Eqref{dev flux} are displayed here.
  }
\label{fig:figure developable 2}
\end{figure*}

As for minimal surfaces, developable surfaces cannot be compact and closed: in the following, we will assume that at some point in space the surface is truncated, even if we are going to ignore boundary effects. In any case, \Eqref{interfaceq} can be written using only Darboux-frame quantities
\be 
\dot{\theta}= \etak \frac{\left(\kappa_n^2+\tau_g^2\right)^2}{\kappa_n^2}+ \frac{\Delta \lambda}{\sigma}\,,
\label{interface dev}
\ee
where, as in the previous Section, we choose $L$ to be a characterizing length-scale of the surface (such as a wave-length). Moreover, the projection onto $\Gamma$ of the Codazzi-Mainardi equation becomes significantly simpler, allowing to explicitly evaluate the second variation of the free energy. For non-fixed $\varphi$, we show in Appendix \ref{app:curves on developable surfaces} that the condition for stability is: 
\be
\dot{\theta}^{2}+ \ddot{\theta} \tan \theta \leq 0 \,.
\label{stability dev}
\ee 
In fact, this relation can never be satisfied for a closed curve in a non-flat region: for the tangent vector direction necessarily spans the full interval $\theta \in [0,2\pi]$, there always exists at least one point on $\Gamma$ where $\tan\theta=0$, i.e. $\bm{T}$ is pointing towards the non-flat principal direction. Being $H\neq0$, then \Eqref{interface dev} implies $\dot{\theta}\neq 0$ for $\etak \neq 0$ and $\Delta\lambda=0$, and thus \Eqref{stability dev} is violated. 

This result shows how the existence of a flat direction renders the stability of finite size domains on developable surfaces impossible, in the case of non-conserved area fractions. In particular, closed and contractible interfaces on cylinders are never stable. This is a similar feature to the one discussed in Sec. \ref{sec:spheres} on domain stability on spheres. The only exception to the above discussion happens if the surface admits points where $H=0$. In this case, geodesics pointing in the flat direction (i.e. curves with $\theta=\pi/2$) have $\nabla_{\bm{N}}H^2=0$ and are thus potentially stable. Geodesic interfaces are generally not closed and this solution correspond to a \textit{striped} phase, where domain boundaries are located at zeroes of the mean curvature. 

This picture changes for conserved area fractions since stability issues are less of a concern: the effect of Lagrange multipliers is to remove zero-mode instabilities. What matters instead is the landscape of equilibrium configurations, which, for non-flat developable surfaces, is highly non-trivial. Thus, for a given value of $\varphi$, we need a general criterion for finding all possible closed interfaces which are local minima of the free energy. In this respect, we find of great help the fact that the interface \Eqref{interface dev} is mathematically identical to the equation of motion of a charged particle moving in a flat plane under the influence of a spatially inhomogeneous axial magnetic field. Upon identifying the arc-length parameter with time and the tangent vector $\bm{T}$ with the particle's planar velocity $\bm{v}$, the geodesic curvature $\kappa_g$ corresponds to nothing more than the acceleration along the planar normal direction $\bm{\tilde{N}}$ (see \fref{fig:figure developable 1}). We then prove in Appendix \ref{app:orbits in a B} that a charged particle moving with constant speed in an axial magnetic field $\bm{B}= B_z \bm{\hat{z}}$ of magnitude
\be
B_z= \etak H^2 + \frac{\Delta \lambda}{\sigma} \,,
\label{Bz dev}
\ee 
will follow a trajectory, determined by the Lorentz force, which coincides with the curve $\Gamma$. Note that the surface's varying curvature is the source of inhomogeneity in the magnetic field, while $\Delta \lambda$ tunes the spatial average of $B_z$. We can thus map the question of finding closed interfaces on a developable surface into the question of finding \textit{closed orbits} of a charged particle moving in a varying magnetic field. In \fref{fig:figure developable 1} we illustrate this analogy for a cylindrical developable surface: for any interface on $\Sigma$ there is a corresponding closed planar trajectory in the $xy$-plane, with the mean curvature being the varying component of the axial magnetic field. 

Note that a generic orbit will not be closed, because a spatially varying magnetic field induces a drift of the center of rotation along a direction perpendicular to both the magnetic field and its gradient: an effect known as \textit{guiding center drift} \cite{Boozer1980}. However, in our set-up we can change the value of the Lagrange multiplier, thus of the average intensity of the field, and tune it in order to obtain a closed orbit. While for constant $\bm{B}$ (i.e. for $\Sigma$ being either a plane or a right cylinder) every trajectory is circular, in general there is only a discrete set of $\Delta\lambda$ values that allow for a closed orbit.

The analogy with electromagnetism nicely carries on also at the functional level:  we can show that the area integral in \Eqref{freeenergy} is simply the magnetic flux $\Phi_{\bm{B}}$ through the area enclosed by the loop $\Gamma$, so that the total free energy is
\be
F = \sigma \ell_\Gamma \pm  \sigma \Phi_{\bm{B}}  (\Sigma_\pm) \,,
\label{dev flux}
\ee
the sign depending on whether the value of $\Delta \lambda$ favours hard or soft domains.

Since the free energy functional is invariant under translations along the flat direction, there is an associated Noether charge, which we identify with the component of the \textit{minimally coupled momentum}
\be
\bm{P} = \bm{v}-\bm{A} \,,
\ee
along the flat direction. Using the charge conservation and the fact that the magnetic flux can be written as the circulation of an electromagnetic potential
\be 
 \Phi_{\bm{B}} (\Sigma_\pm) = \mp \int_\Gamma \D s\; \bm{v} \cdot \bm{A} \,,
\ee
we are able to write the free energy as a single line integral over $\Gamma$, namely:
\be 
F = \sigma \int_{\Gamma} \D s\,v_x^2\;,	
\label{freeenergy dev flux}
\ee
where $v_x$ is the component of the velocity $\bm{v}$ along the curved direction, see Appendix \ref{app:orbits in a B} for more details. This expression is of great help in numerical applications.
 
In \fref{fig:figure developable 2} we show how this applies to the wave-like cylindrical developable surface of \fref{fig:figure developable 1}. For different values of $\etak$, we found the initial conditions (i.e. the value of $x(0)=x_0$ where $\bm{v}$ points in the $x$ direction) and the correct $\Delta \lambda$ such that \Eqref{dev flux} is minimized and the area of the enclosed domain is fixed to a given value. To evaluate the free energy we used \Eqref{freeenergy dev flux}. We do this for both soft domains (curves with $\kappa_g>0$, left panel of \fref{fig:figure developable 2}) and for hard domains (curves with $\kappa_g<0$, right panel). By increasing the values of $\etak$ the phase domain tends to become more and more elongated with its center lying in regions of maximal curvature (either the valleys or ridges of the sinusoidal profile of \fref{fig:figure developable 1}). If the curvature is strong enough the domain develops concavities. 

While $\etak$ is a material property (eventually fixed by the types of lipids involved in the phase separation), both the height and periodicity of $\Sigma$ are movable parameters: in principle it should be possible to scale the shape of the surface so that each of the domains in \fref{fig:figure developable 2} is obtained. Conversely, by observing a specific domain shape for a given geometry, it should be possible to find the value of $\etak$ even in the case of a fixed area fraction $\varphi$.

As final remark of this Section, note that even though in \fref{fig:figure developable 1} and \fref{fig:figure developable 2} we used a cylindrical surface, our magnetic analogy applies equally well to every class of surfaces with $K=0$, i.e. also to a conical or a tangent-developable $\Sigma$.

\section{Discussion}

\label{sec:discussion}

In this article we report an exhaustive theoretical investigation of the equilibrium configurations of binary mixtures on curved substrates. Our main motivation stems from the physics of lipid bilayers supported by solid substrates, but most of our results are generally valid and apply, upon adjusting the relevant material parameters, to arbitrary two-dimensional binary mixtures freely diffusing across non-flat surfaces. A versatile experimental realization of this paradigm, that we recently introduced in \cite{Rinaldin2018} and here refer to as scaffolded lipid vesicles (SLVs), consists of arbitrarily shaped colloidal particles coated with mixtures of saturated and unsaturated lipids. In this example, a small percentage of the lipids are anchored to a silica substrate by mean of polyethylene glycol (PEG) molecules, while the bilayer preserves lateral mobility for the majority of its components. At room temperature, the lipids separate in two phases, LO and LD, having different internal order and bending moduli. 

As it was already predicted in a classic paper by J\"ulicher and Lipowsky \cite{Julicher1993}, and reviewed by us in Sec. \ref{sec:curved}, the difference in stiffness between the two lipid phases introduces a coupling mechanism between the chemical composition of the lipids and substrate curvature, whose primary effect is to pin the stiffer phase in regions of low curvature, at the expenses of the softer phase. When the bending energy difference is sufficiently large to overcome interfacial tension, this mechanism might lead to the formation of multiple finite domains of one phase. As we explain in Sec. \ref{sec:curved}, however, the existence of multiple domains alone does not imply a direct coupling between composition and curvature, as interfaces on a curved surface can be simultaneously curved and length-minimizing. 

In this work we have highlighted with special care the role of the area fraction $\varphi$ (i.e. the percentage of the total available area covered by either one of the two phases) and demonstrated how this dramatically affects interfacial stability. Upon minimizing the J\"ulicher-Lipowsky free energy on a generic curved surface, we derived a curved-space analogue of the Young-Laplace law, \Eqref{interfaceq}, from which we could identify three fundamental scenarios. In the absence of direct coupling with the curvature, interfaces lie along geodesics (for non-conserved area fraction), or lines of constant geodesic curvature (for conserved area-fraction). A direct coupling with the curvature introduces an additional, space-dependent, Laplace pressure at the interface, proportional to the difference between the bending moduli and to the local mean squared and Gaussian curvatures. This causes the interface to deviate from the local geodesics and to become more and more curved the larger the difference in stiffness between the two phases is (Sec. \ref{sec:local expansion}). In all these cases, negative Gaussian curvature enhances the stability of the interface, since deviations from minimal shapes are necessarily penalized.
 
In Sec. \ref{sec:examples} we have then restricted our analysis to specific classes of surfaces of both practical and conceptual interest. In the case of spherical substrates (Sec. \ref{sec:spheres}), we showed that, for non-conserved area fractions, interfaces are always unstable and equilibrium is achieved upon expelling the stiffer phase from the spherical substrate. For conserved area fractions, on the other hand, a stable equilibrium configuration consists of a single circular interface, regardless of the difference in the bending moduli. In Sec. \ref{sec:axisymmetric} we considered axisymmetric surfaces, whose geometry is completely determined by the shape of an axial cross-section. Thanks to their simplicity, this class of surfaces has played a special role in the literature \cite{deuling1976curvature,jenkins1977static,Julicher1996,Baumgart2005} and represents the only case where analytical progress can be made even in the general problem, where both the shape of the membrane and geometry of the interfaces are allowed to change. Furthermore, axisymmetric membranes have been experimentally investigated, both in the context of GUVs \cite{Baumgart2003} and by us in SLVs \cite{Rinaldin2018}. Whereas nearly every theoretical work on binary axisymmetric membranes is built upon the assumption that interfaces on axisymmetric surfaces are themselves axisymmetric, in Sec. \ref{sec:axisymmetric} we show that several non-axisymmetric interfaces can exist for both conserved and non-conserved area fractions. In the case of axisymmetric interfaces, we mapped out a complete phase diagram in terms of the area fraction $\varphi$ and dimensionless number $\etak$ and $\etakb$ expressing the relative contribution of bending and interfacial tension to the total energy. In the case of minimal surfaces (Sec. \ref{sec:minimal}), the stability of the interface depends exclusively on the surface Gaussian curvature. Using the Weierstrass-Enneper (WE) parametrization, we introduced a generic shape equation, well suited for numerical analysis, and used it to investigate interfaces on triply periodic minimal surfaces, with possible applications to complex lipid assemblies \cite{Hyde1984,Seddon2000,Kaasgaard2006,Demurtas2015}. In Sec. \ref{sec:developable}, finally, we considered developable surfaces and show that, as spherical substrates, they cannot support stable closed interfaces for vanishing Laplace pressure. Furthermore, taking advantage of a fascinating analogy with electrodynamics, we derived an extremely concise expression for the system free energy that could provide a valuable tool in combination with experiments on supported lipid bilayers. 

Whereas the study of phase separation in lipid bilayers is a classic subject in membrane physics, the recent experimental and theoretical developments, including those reported here and in Ref. \cite{Rinaldin2018}, offer a promising route for further progress. Both {\em in vivo} and in artificial lipid membranes, for instance, the curvature of the bilayer can be locally adjusted by incorporating asymmetric lipid molecules (see e.g. \cite{Kuzmin2005}) or curvature-generating proteins (see e.g. \cite{Rautu2015}) into either one of the leaflets. Theoretically, this amounts to add a spontaneous curvature in the Canham-Helfrich free energy, such that the $H^{2}$ term in \Eqref{freeenergy} becomes $(H-H_{0})^{2}$, with $H_{0}$ a constant parameter. This term manifestly breaks the symmetry between the opposite sides of a membrane, reflecting the fact that curvature-generating inclusions bind only to one of the leaflets. In the future, we plan to extend the approaches proposed here to include spontaneous curvature and explore how the latter can conspire with the substrate curvature to control the spatial organization of lipid domains. Finally, the special cases where the J\"ulicher-Lipowsky free energy can be cast in the form of a line-integral (e.g. Secs. \ref{sec:minimal} and \ref{sec:developable}) are especially well-suited to investigate the role of fluctuations. Ideally, one could envision a new generation of substrates whose geometry is specifically designed to enhance the amplitude of certain modes, with the two-fold purpose of obtaining more accurate estimates of the material parameters and gain insight into the complex physics of interfaces in curved geometries. 

\paragraph*{Acknowledgements} 
This work was supported by the Netherlands Organisation for Scientific Research (NWO/OCW), as part of the Frontiers of Nanoscience program (MR), VENI grant 680-47-431 (DJK) and the VIDI scheme (LG,PF). PF wants to thank \'Alvaro V\'eliz-Osorio for useful comments on an earlier version of the manuscript.

\appendix 

\section{Geometry of curves and surfaces}

\label{app:notation}

In this Appendix, we review some essential concepts of differential geometry of surfaces and embedded curves on surfaces and we clarify our notation. As a general rule, we denote tensor fields on $\mathbb{R}^3$ using Greek indices $\lambda,\nu,\rho$ while tensors on $\Sigma$ have Latin indices $i,j,k$. The arc-length parameter of $\Gamma$ is denoted by $s$. 

\subsection{Surfaces}

\label{app:surfaces}

A generic surface immersed in three-dimensional Euclidean space can be described by an explicit parametrization $\bm{r}=\bm{r}(\sigma^{1},\sigma^{2})$, where ${\sigma^{1},
\sigma^{2}}$ are local coordinates on $\Sigma$. A basis for the tangent space of $\Sigma$, which we call $T\Sigma$, is given by the vector fields 
\be
\bm{t}_{i} = \partial_{i} \bm{r}\;,
\label{tangent sigma}
\ee
where $\partial_i$ is the derivative with respect to $\sigma^i$. Since the space is three-dimensional, there exists  a unique, up to orientation, unit norm vector field $\bm{n}$ which is orthogonal to the tangent plane at every point of $\Sigma$. The triplet $\{\bm{t}_{1},\bm{t}_{2},\bm{n}\}$ define an oriented orthonormal frame of $T\mathbb{R}^3$ at any point on $\Sigma$. 
The induced metric on the surface is then
\be 
h_{ij} = \bm{t}_{i}\cdot\bm{t}_{j}\;,
\label{induced metric}
\ee
which is a symmetric tensor. From $h$ we can construct intrinsic connections
\be 
\Gamma_{ij}^k = \bm{t}^{k}\cdot \partial_i \bm{t}_{j} \,,
\label{Christoffels}
\ee
which allow to define the covariant derivative $\nabla_i$ acting on surface tensors. In particular, this connection is by construction metric-compatible, i.e. $\nabla_i h_{jk} = 0$. The Riemann tensor $\mathcal{R}_{ijkl}$ in two-dimension has always one independent component, and its tensorial structure is completely fixed by the induced metric
\be
\mathcal{R}_{ijkl} = \frac{\mathcal{R}}{2} \left( h_{ik} h_{jl} - h_{il} h_{jk} \right)  \,,
\label{two d R}
\ee
where $\mathcal{R}$ is the intrinsic Ricci curvature  of $\Sigma$. The extrinsic curvature tensor is defined as
\be 
K_{ij} = \bm{t}_i\cdot \partial_j \bm{n} \,.
\label{Kij}
\ee
Both $\Gamma_{ij}^k$ and $K_{ij}$ are symmetric for exchange of the indices $ij$. Note that $K_{ij}$ sometimes defined in the literature with an opposite sign than \Eqref{Kij}. This is a matter of conventions which carries no geometrical meaning because it can always be compensated by a change of normal field orientation: $\bm{n}\rightarrow -\bm{n}$. 

From the metric and extrinsic curvature tensors one can extract two geometric invariants, the \textit{mean} and \textit{Gaussian} curvatures, defined as:
\begin{subequations}\label{eq:curvatures}
\begin{align}
H &= \frac{1}{2} h^{ij} K_{ij} \,,\\
K &= \det (h^{ik}K_{jk}) \,.
\end{align}
\end{subequations}
The eigenvalues of the matrix $K^{i}_{j}=h^{ik}K_{jk}$ are the principal curvatures of the surface, which we denote by $\kappa_i$. Similarly, the eigenvectors define two vector fields on $\Sigma$, called principal directions, which we denote by $\bm{k}_i$. Such vector fields are well defined as long as the principal curvatures are non-degenerate; points where $\kappa_1=\kappa_2$ are known as umbilical.

It is a well known fact that a surface in $\mathbb{R}^3$ is defined, up to Euclidean isometries, if $h_{ij}$, $K_{ij}$ and $\mathcal{R}$ are given (see e.g. \cite{spivak1999vol3}). However, these quantities cannot be arbitrarily chosen but need to satisfy a set of integrability conditions. In the particular case of surfaces, these conditions are known as the \textit{Gauss} and the \textit{Codazzi-Mainardi} relations. 
The Gauss relation in 2D takes the remarkably simple form
\be
\mathcal{R} = 2 K \,,
\label{theorema egregium}
\ee
which is known as Gauss' \textit{Theorema Egregium}. The fact that the Ricci intrinsic curvature is directly proportional to the Gaussian curvature is the reason we did not need to include a term proportional to $\mathcal{R}$ in \Eqref{freeenergy}.
Furthermore, the Codazzi-Mainardi relations
\be 
\nabla_{i} K_{jk} - \nabla_{j} K_{ik} = 0 \,,
\label{codazzi mainardi}
\ee
constrain how the extrinsic curvature is allowed to vary along the surface. 

The induced metric allows to define an invariant measure on $\Sigma$, which we denote by
\be
\D A = \D \sigma^1 \D \sigma^2 \sqrt{\det h} \,,
\label{dA}
\ee
so that we can perform integrals over the surface. 
By means of \Eqref{dA}, it is possible to prove that the integration over $\Sigma$ of \Eqref{theorema egregium} leads to the \textit{Gauss-Bonnet} theorem for compact surfaces without boundaries
\be
\int_\Sigma \D A  \, K = 2 \pi \chi  \,,
\label{gauss bonnet sigma}
\ee
where $\chi=2(1-g)$ is the Euler characteristic of a $\Sigma$ with genus $g$.

\subsection{Curves}

\label{app:curves}

We now consider the embedding of the interface, i.e. of the curve $\Gamma$, into the two-dimensional surface $\Sigma$.
In general, we can always construct an explicit parametrization of $\Gamma$ by defining two functions $\sigma^i(s)$ (we remind that $\sigma^i$ are the generic coordinates on $\Sigma$). The \textit{intrinsic} tangent vector to the curve is 
\be 
T^i =\frac{\D\sigma^i}{\D s} = \dot{\sigma}^i \,,
\ee
where $s$ is the parameter that spans throughout the curve. Since the intrinsic geometry is trivial, we can always fix the normalization of the tangent vector by a reparametrization of $s$. Fixing this norm to be equal to one gives the arc-length condition 
\be
T^i T^j h_{ij} = 1 \,. 
\label{arclength}
\ee
We furthermore define $N^i$ to be the two-vector normal to $T^i$ and pointing into $\Sigma_+$ domains.
Notice that condition \Eqref{arclength} can be true only along $\Gamma$, since in general it is impossible to maintain \Eqref{arclength} true along the normal direction: 
\be 
N^i \nabla_i (T_j T^j)  \neq 0 \,.
\ee 
From now on, we will always assume that the curve is measured by its arc-length. 
The rate of variation of $N^i$ when moving along $\Gamma$ is captured by the \textit{geodesic curvature} 
\be 
\kappa_g = T^i T^j \nabla_i N_j \,,
\label{geodesic curvature}
\ee
which measures the departure of $\Gamma$ from being a geodesic and, as for the extrinsic curvature, its overall sign is matter of pure convention. With our definitions we choose $\kappa_g$ to be positive whenever $\Sigma_-$ is a convex domain. It is possible to prove that the Gauss-Bonnet theorem \Eqref{gauss bonnet sigma} generalizes to the case of surfaces with boundary as	
\be 
\int_{\Sigma_\pm} \D A  \, K 
\mp
\int_\Gamma \D s \, \kappa_g
=
2 \pi \chi(\Sigma_\pm)  \,.
\label{gauss bonnet}
\ee

We can also see the curve $\Gamma$ as directly embedded in the real three-dimensional space. In this case the curve posses one tangential and two normal vectors. We can promote both $T^i$ and $N^i$ to vector fields on the tangent space of $\mathbb{R}^3$ via push-forward
\be
\bm{T} = T^i \bm{t}_i \,, \quad
\bm{N} = N^i \bm{t}_i \,.
\ee
It is precisely these vectors, and not $T^i$ or $N^i$, that are depicted in \fref{fig:figure 1}.
The co-moving frame with basis vectors $\{\bm{T},\bm{N},\bm{n}\}$ is known as \textit{material} or \textit{Darboux frame} of $\Gamma$.

In general, the shape of a curve in three-dimensional space is captured by three quantities, two curvatures and one torsion, of which two only are independent because of the freedom in choosing the orientation of the normal frame along the curve. In the Darboux frame, one of these two curvatures is provided by $\kappa_g$, while the other, known as \textit{normal curvature} is defined as the rate of rotation of $\bm{n}$ projected onto $\bm{T}$ while moving along $\Gamma$. It can be easily proven (see, for example, \cite{envirobias}) that $\kappa_n$ is equal to the projection along $T^i$ of the extrinsic curvature evaluated on $\Gamma$, namely
\be
\kappa_n 
=
\bm{T} \cdot \dot{\bm{n}}
=
T^i T^j K_{ij} \,.
\label{normal curvature}
\ee
Furthermore, the material frame has a \textit{geodesic torsion}, which - as it will be clear later - measures the deviation of $\bm{T}$ from a principal direction. It is defined as the rate of rotation of $\bm{N}$ in the direction $\bm{n}$ while moving along $\Gamma$. Similarly to \Eqref{normal curvature}, it is easy to show that it is equal to the projection of the extrinsic curvature onto the curve frame
\be
\tau_g 
=
\bm{n} \cdot \dot{\bm{N}}
= 
- N^i T^j K_{ij} \,.
\label{geodesic torsion}
\ee

To characterize completely the projection of the extrinsic curvature onto the Darboux frame a further quantity is needed, which measures the change of direction of $\bm{n}$ when moving on $\Sigma$ but away from $\Gamma$ itself. This quantity has not a generally accepted name, and strictly speaking is not part of the Darboux frame: it does not describe a property of the curve but rather expresses how the surface bends in the normal direction, using the curve's frame. We call it $\Theta$ (for example it was called $h_p$ in \cite{Elliott2010}) and define it to be
\be
\Theta = N^i N^j K_{ij} \,. 
\ee
With this notation, we see we can decompose the induced metric on $\Gamma$  as 
\be 
\left. h_{ij}\right|_\Gamma = T_i T_j + N_i N_j \,,
\ee
and the extrinsic curvature as
\be 
\left. K_{ij}\right|_\Gamma = \kappa_n T_i T_j -\tau_g \left( N_i T_j + N_j T_i \right) + \Theta N_i N_j \,.
\ee
The four scalar functions $\kappa_g$, $\kappa_n$, $\tau_g$ and $\Theta$ completely characterize the curve in three dimensions and its relation to the surface. They are not completely independent since the torsion the extrinsic curvature has to satisfy the Codazzi-Mainardi relations \Eqref{codazzi mainardi}. 

At last, we can use the above results to express the Gaussian and mean curvatures evaluated on the curve, which enter Eqs. \eqref{interfaceq} and \eqref{stability}, in terms of Darboux frame quantities
\begin{subequations}
\begin{align}
\left. H \right|_\Gamma &= \frac{1}{2} \left( \kappa_n + \Theta \right) \,, 
\label{H gamma}\\
\left. K \right|_\Gamma &= \kappa_n \Theta - \tau_g^2 \,.
\label{KG gamma}
\end{align}
\end{subequations}

Above, the symbol $\vert_\Gamma$ next to any quantity indicates that the expression should be evaluated along the curve, rather than on a generic point on the surface. In the main text, and in the following Sections, we will drop this notation for the sake of readability.

\subsubsection{From Darboux to other frames}

Sometimes it will turn out to be useful to express the curve's geometric invariants using other frames rather than Darboux. As we mentioned in Appendix \ref{app:surfaces}, the tangent space of surfaces without umbilical points can be described by the span of the eigenvectors of the extrinsic curvature. Since in the proximity of $\Gamma$ also the orthonormal pair $\{ \bm{T}, \bm{N}\}$ forms a basis for $T\Sigma$, there exist a local $SO(2)$ rotation matrix that links these two frames, since the orientation of the principal frame can be arbitrarily chosen. The two bases are related by the transformation
\begin{subequations}
\begin{align}
\bm{T} &=
\cos \theta \bm{k}_1 
+ 
\sin \theta \bm{k}_2 \,,
\\
\bm{N} &=
-\sin \theta \bm{k}_1 
+
\cos \theta \bm{k}_2 \,,
\end{align}
\end{subequations}
where $\theta = \theta(s)$ is the local angle between the two frames. With this choice, it is easy to show that 
\begin{subequations}
\begin{align}
\kappa_n &= \kappa_1  \cos^2 \theta \; +  \kappa_2  \sin^2 \theta \, \,,
\label{kappantheta} \\
\tau_g &= \left( \kappa_1- \kappa_2 \right) \sin \theta \; \cos \theta\,,
\label{taugtheta}\\
\Theta &= \kappa_1  \sin^2 \theta \; +  \kappa_2  \cos^2 \theta  \,.
\cos \theta \bm{k}_2 \,,
\end{align}
\end{subequations}
The first of these equations is known as the \textit{Euler formula}. These expressions make evident that a curve following a principal direction, say $\bm{k}_1$, has normal curvature equal to $\kappa_1$, vanishing geodesic torsion and $\Theta=\kappa_2$, showing how $\Theta$ encodes the information about how the surface bends away from the curve.

It is also possible to derive a similar expression for the geodesic curvature \Eqref{geodesic curvature}, which transforms under a frame rotation as
\be
\kappa_g
=
\dot{\theta}+ \cos \theta \; \kappa_g(\bm{k}_1) + \sin \theta \; \kappa_g(\bm{k}_2) \,, 
\label{kg theta k1 k2}
\ee
an expression which sometimes is known as \textit{Liouville's formula} \cite{DoCarmo1976}, but can be seen as just an explicit representation of the non-tensorial nature of Christoffel symbols. Here $\kappa_g(\bm{k}_1),\;\kappa_g(\bm{k}_2)$ are the geodesic curvatures of the lines of curvature evaluated on $\Gamma$.

As a final remark, remember that one can choose the co-moving frame of $\Gamma$ in such a way that the total curvature is captured by a single normal vector: such a frame is known as \textit{Frenet-Serret} (FS), whose geometric invariants are the total curvature $k_{FS} \geq 0$ and the Frenet-Serret torsion $\tau_{FS}$. The map between Frenet-Serret and Darboux frames is given by the relations
\begin{subequations}
\begin{align}
\kappa_{FS} &= \sqrt{\kappa_n^2 + \kappa_g^2}  \,, \\
\tau_{FS} &= \tau_g + \frac{\dot{\kappa}_n \kappa_g - \dot{\kappa}_g \kappa_n}{\kappa_n^2 + \kappa_g^2} \,.
\end{align}
\end{subequations}
The functions $\tau_{FS}$, rather than measuring departure from principal directions, is vanishing for planar curves only.

Using the FS frame usually simplifies greatly the description of curves embedded in three-dimensional Euclidean space. However, if $\Gamma$ is constrained to lie on a particular submanifold, it becomes counter-intuitive. For this reason we will never use it in the following. Note that for geodesics we have $\kappa_{FS} = |\kappa_n|$ and $\tau_{FS} = \tau_{g}$.

\subsection{Local expansions}

\label{app:local expansion}

In Section \ref{sec:local expansion} of the main text we show how to solve the interface equation in a neighbourhood of a point of $\Gamma$. If $\Sigma$ has no degenerate saddle points, it is always possible to locally express the surface as a quadric of the type \Eqref{monge quadric}. We pick as $x,y$ axes the two local principal directions at the generic point $P \in \Sigma$:
\be
\bm{\hat{x}}=\bm{k}_1(P) \,, \quad \bm{\hat{y}}=\bm{k}_2(P) \,,
\ee 
while the $z$ axis is given by the surface normal $\bm{\hat{z}}=\bm{n}(P)$. We choose the coordinates $\{x,y,z\}$ so that the point $P$ is at the origin of the Cartesian axes.
The induced metric \Eqref{induced metric} is then
\be
h_{ij} = 
\left(
\begin{array}{c c}
1 + \kappa_1^2 x^2 & \kappa_1 \kappa_2 x y \\
\kappa_1 \kappa_2 x y &  1+\kappa_2^2  y^2
\end{array}
\right)  \,,
\label{local hij}
\ee
from which we deduce the extrinsic curvature tensor \Eqref{Kij} 
\be
K_{ij} = 
\frac{1}{\sqrt{1+ \kappa_1^2 x^2 + \kappa_2^2 y^2}}
\left(
\begin{array}{c c}
\kappa_1 & 0 \\
0 &  \kappa_2
\end{array}
\right)  \,.
\label{local Kij}
\ee
Using Eq. (\ref{eq:curvatures}a) and expanding around $\{x,y\}\approx \{0,0\}$, we find:
\be
H^2 =
H_{0}^{2}
\left(
1
-
\delta_1 \kappa_1^2 x^2 
- 
\delta_2 \kappa_2^2 y^2 
\right) \,,
\label{H2epsilon}
\ee
where $H_{0}=(\kappa_{1}+\kappa_{2})/2$ is the mean curvature at the origin and $\delta_i = 1+2\kappa_i /(\kappa_1+\kappa_2)$. Similarly, from the determinant of $K_i^{\;j}$ we get the expansion
\be
K =
K_{0}
\left[
1 
-
2 \left( \kappa_1^2 x^2 + \kappa_2^2 y^2\right)
\right] \,,
\label{KGepsilon}
\ee
with $K_{0}=\kappa_{1}\kappa_{2}$ the Gaussian curvature at the origin. 

We can specify a curve on $\Sigma$ by defining two functions $\{x(s),y(s)\}$. The arc-length condition is satisfied by parametrizing the tangent vector as
\be 
\bm{T}
=
\frac{
\cos \theta \bm{\hat{x}}
+
\sin \theta \bm{\hat{y}}
}{\sqrt{1+(\kappa_1 x \cos \theta +\kappa_2 y \sin \theta)^2}}
\,,
\label{local tangent}
\ee
with $\theta=\theta(s)$. The above definition establishes a first-order differential relation between $\theta$, $\dot{x}$ and $\dot{y}$. By using the definition \Eqref{geodesic curvature} along with the covariant derivative compatible with the metric \Eqref{local hij}, we can compute the geodesic curvature of $\Gamma$ in a neighbourhood of the point $\{x,y\}=\{0,0\}$ for small $s$.
The expansion of $\kappa_g$ up to second order in arc-length gives
\bea
\begin{aligned}
\kappa_g =& \theta^{(1)}_ 0 + s \left( \theta^{(2)}_0  -\kappa_{n0} \tau_{g0}\right) + 
\\ & + 
\frac{1}{2}\left(\left(K_0- 3\kappa_{n0}^2 + 6 \tau_{g0}^2 \right) \theta^{(1)}_ 0 +\theta^{(3)}_ 0 \right) s^2 + \dots
 \,,
\end{aligned}
\label{kg local 2}
\eea
with $\kappa_{n0}=\kappa_n(0)$ and $\tau_{g0}=\tau_g(0)$.
Truncating this expression at order $s$ gives \Eqref{kg local}. Similarly, one can take Eqs. \eqref{H2epsilon} and \eqref{KGepsilon} and, by using the small $s$ expansion of \Eqref{local tangent}, obtain Eqs. \eqref{H2 local} and \eqref{K local}.

\section{Variational calculus with curves}

\label{app:variational}

In this Appendix we derive \Eqref{interfaceq} and \Eqref{stability} by calculating the first and second variation of \Eqref{freeenergy}. For simplicity, we assume that $\Gamma$ consists of a single curve, but all results generalize to more complicated interface topologies. 
The only continuous degree of freedom in our problem is the position of $\Gamma$: we need to study the response of the free energy under an infinitesimal shift $\Gamma \to \Gamma + \delta \Gamma$. The deformation $\delta \Gamma$ is forced to lie on $\Sigma$ because we do not allow the membrane to change its shape. Since $\Gamma$ is closed and the free energy does not depend on the curve parametrization, any tangential deformation can always be adsorbed in a redefinition of the curve parameter. The most generic non trivial deformation is thus captured by a purely normal shift, which we can express in the explicit form as
\be 
\bm{X} \to \bm{X} +  \epsilon_{\bm{N}} \bm{N}\;,
\label{normal deformation}
\ee
where $\epsilon_{\bm{N}}=\epsilon_{\bm{N}}(s)$ is the deformation parameter, which we assume to be small enough so that every result in the following has to be intended as an expansion at first order in $\epsilon_{\bm{N}}$. Recall that we define the direction of $\bm{N}$ to point in the $\Sigma_+$ domains, see \fref{fig:figure 1}. Given \Eqref{normal deformation}, one can compute variations of geometrical quantities. For instance, the tangential and normal unit vectors change as
\begin{subequations}
\begin{align}
\delta \bm{T} &= \dot{\epsilon}_{\bm{N}} \bm{N} \,, \\
\delta \bm{N} &= -\dot{\epsilon}_{\bm{N}} \bm{N} \,,
\end{align}
\end{subequations}
Note that if we had not unit-normalized $\bm{T}$, its variation would have contained a further tangential term proportional to $\kappa_g$.
It is a bit more complicated to derive variations for other quantities (see e.g. Ref. \cite{langer1984}), but one  can prove that the geodesic curvature changes as
\be
\delta \kappa_g = -\epsilon_{\bm{N}} \left( \kappa_g^2+  K \right) - \ddot{\epsilon}_N \,.
\label{var kg}
\ee
Furthermore, it is possible to prove that the variation of the normal curvature and of the geodesic torsion are  respectively \cite{Fonda:2016ine}
\begin{subequations}
\begin{align}
\delta \kappa_n &= \epsilon_{\bm{N}} \left( \kappa_g (\Theta - \kappa_n)- \dot{\tau}_g \right) - 2 \dot{\epsilon}_{\bm{N}} \tau_g  \,,
\label{var kn} \\
\delta \tau_g &=\epsilon_{\bm{N}} \left(\dot{\Theta}-2 \kappa_g \tau_g \right) + \dot{\epsilon}_{\bm{N}} \left(\Theta-\kappa_n  \right)   \,. 
\label{var taug}
\end{align}
\label{var kn taug}
\end{subequations}
Using these results one can compute variations of the terms appearing in \Eqref{freeenergy}.

\subsection{The first variation}

\label{app:first variation}

The first normal variation of the curve length is proportional to the integral of the geodesic curvature
\be
\delta \int_\Gamma \D s 
=
\int_\Gamma \D s \, \epsilon_{\bm{N}} \kappa_g \,. 
\label{var ell}
\ee
The free energy contains terms involving area integrals, whose domains of integration $\Sigma_\pm$ are bounded by $\Gamma$: the shift $\Gamma\to \Gamma+\delta \Gamma$ induces a change in the extension of the domain of integration, focused near the boundary. Therefore the response to such shift can be expressed in terms of boundary line integrals. To make precise statements consider an arbitrary function $f=f(\sigma^i)$ defined on $\Sigma$. The first normal variation of its integral over $\Sigma_\pm$ is given by
\be
\delta \int_{\Sigma_\pm} d A \, f = \mp \int_\Gamma \D s \, \epsilon_{\bm{N}} f \,,
\label{var f area}
\ee
where the sign in front of the variation follows from our convention for $\bm{N}$. Eqs. \eqref{var ell} and \eqref{var f area} are all what is needed in order to derive the interface equation, once we replace the function $f$ with either $H^2$ or $K$. Namely, we have
\be 
\delta F =
\int_\Gamma
ds \, \epsilon_{\bm{N}}
\left( 
\sigma \kappa_g -
\Delta k H^2
-
\Delta \bar{k} K	 
-
\Delta \lambda \right) \,. 
\label{var F}
\ee
Since $\epsilon_{\bm{N}}$ is small but arbitrary, by requiring $\delta F = 0$ we obtain \Eqref{interfaceq}.
As a check of our methods, let us consider the Gaussian bending terms of \Eqref{freeenergy} and let us rewrite them by means of the Gauss-Bonnet theorem \Eqref{gauss bonnet}
\be 
\sum_{\alpha=\pm}
\int_{\Sigma_\alpha} \D A \,
\bar{k}_\alpha
K
=
\Delta \bar{k} \int_\Gamma \D s \, \kappa_g
+
2\pi \bar{\chi} \,,
\label{GB theo kg}
\ee
where $\bar{\chi} = \bar{k}_+ \chi_+ +  \bar{k}_- \chi_{-}$ is a topological term. Taking the normal variation of this expression and assuming that $\epsilon_{\bm{N}}$ is small enough not to change the topology of $\Sigma_\pm$ we can use \Eqref{var kg} to obtain
\be 
\delta \int_\Gamma \D s \, \kappa_g
=
- \int_\Gamma \D s \, \epsilon_{\bm{N}} K \,,
\ee
correctly reproducing the $\Delta \bar{k}$ term in \Eqref{var F}.

\subsection{The second variation}
\label{app:second variation}

For a given surface, \Eqref{interfaceq} has in general many non-equivalent solutions and it is therefore of utmost importance to distinguish stable from unstable configurations. This is obtained by studying the second variation of the geometric functional under consideration, which essentially correspond to study terms of order $\epsilon_{\bm{N}}^2$ in the expansion of $F$ after the deformation \Eqref{normal deformation}.

As it is customary with standard derivatives, the second variation can be computed as the variation of the first variation, evaluated on the original $\Gamma$: therefore, all we need to do is to compute the variations of \Eqref{var ell} and \Eqref{var f area}. For the former, we have
\be 
\delta^{(2)} \int_\Gamma \D s
=
\int_\Gamma \D s \left[ (\dot{\epsilon}_{\bm{N}})^2 - \epsilon_{\bm{N}}^2   K \right] \,,
\label{eq:second_variation_1}
\ee
where we performed an integration by parts. For the latter we have
\be
\delta^{(2)} \int_{\Sigma_\pm} \D A \, f 
=
\mp \int_\Gamma \D s \, \epsilon_{\bm{N}}^2 \left( \kappa_g f + \nabla_{\bm{N}} f \right) \,,
\label{eq:second_variation_2}
\ee
where $\nabla_{\bm{N}} = N^i \nabla_i$ is the directional covariant derivative along the curve's normal $\bm{N}$. Using Eqs. \eqref{eq:second_variation_1} and \eqref{eq:second_variation_2}, it is then possible to take the variation of \Eqref{var F} and evaluate it on a solution satisfying \Eqref{interfaceq}, obtaining
\begin{widetext}
\be 
\delta^{(2)} F
=
\int_\Gamma \D s \left[
\sigma (\dot{\epsilon}_{\bm{N}})^2
-
\epsilon_{\bm{N}}^2 
\left(
\sigma K
+
\sigma 
\kappa_g^2
+
\Delta k \nabla_{\bm{N}} H^2
+
\Delta \bar{k} \nabla_{\bm{N}} K 
\right)
\right] \,.
\label{var var F}
\ee
\end{widetext}
The first term in this expression is always positive; if we allow fluctuations of the interface to be arbitrary, then requiring minimality, i.e. $\delta^{(2)} F  > 0$, implies the condition \Eqref{stability}. 

It is however not always the case that the interface can fluctuate without constraints. If the area fraction $\varphi$ is fixed, not every choice of $\epsilon_{\bm{N}}$ is permitted: the areas occupied by the two phases, $A_+$ and $A_-$, are not allowed to change and by requiring them to have zero total variation one obtains
\be 
\delta A_\pm = \mp \int_\Gamma \D s \, \epsilon_{\bm{N}}  = 0\,,
\label{var Apm}
\ee
which clearly constrains the choice of the function $\epsilon_{\bm{N}}$. To gain more insight and understand the implications on the positivity of \Eqref{var var F}, it is convenient to decompose the deformation parameter in its Fourier modes.  It is our assumption that $\Gamma$ consists of a single closed curve, but the generalization to multiple interfaces is straightforward. We then have
\be 
\epsilon_{\bm{N}}  = \sum_{n\in \mathbb{Z}} \epsilon_n e^{ i \omega_n s }  \,,
\label{epsilon Fourier}
\ee
where $\epsilon_n$ are Fourier coefficients, $\omega_n =2\pi n/\ell_\Gamma$ and $\ell_\Gamma$ is the total length of the interface. The reality of $\epsilon_{\bm{N}}$ implies that $\epsilon_{-n}^* = \epsilon_n$. With this notation, condition \Eqref{var Apm} tells us that the deformation has vanishing zero mode
\be
\epsilon_0 = 0 \,.
\ee
Let us further define 
\be 
Q = 
\sigma 
\left( K 
+
\kappa_g^2
\right)
+
\Delta k \nabla_{\bm{N}} H^2 
+
\Delta \bar{k} \nabla_{\bm{N}} K  \,,
\label{Q def}
\ee
where $Q=Q(s)$ is a real function on $\Gamma$ which can be decomposed in its Fourier modes $Q_n$ as in \Eqref{epsilon Fourier}. Then \Eqref{var var F} can be rewritten as
\be 
\delta^{(2)} F
=
\sigma \frac{8 \pi^2}{\ell_\Gamma}
\sum_{n>0}
n^2 |\epsilon_n|^2 
-
\sum_{n,m\neq 0}
\epsilon_n \epsilon_{m}^*
Q_{n-m}
\,.
\label{var var F constrained}
\ee
In cases where $Q$ does not depend on the arc-length parameter, we can write $Q_{n-m}=Q_0 \delta_{n,m}$ and the above expression further simplifies to
\be 
\delta^{(2)} F
=
2
\sum_{n>0}
|\epsilon_n|^2 
\left(
\sigma \frac{4 \pi^2}{\ell_\Gamma}
n^2 
- \ell_\Gamma Q_0
\right)
\,,
\label{var var F constrained constant g}
\ee
so that the stability condition becomes, for systems satisfying \Eqref{var Apm}, the inequality
\be 
Q_0 < \sigma \frac{4 \pi^2}{\ell_\Gamma^2} \,.
\ee

\subsection{The role of interface topology}

\label{app:gamma topology}

When $\Gamma$ consists of multiple simple curves then some extra care is required when deriving the equation of motion from \Eqref{freeenergy}. In general $\Sigma$ is partitioned into a collection of $N_+ + N_-$ domains, where $N_\pm$ count the number of domains of a given phase. We can then decompose each single phase domain into a union over connected components
\be
\Sigma_\pm = \bigcup_{l=1}^{N_\pm} \Sigma_\pm^{(l)} \,.
\ee
Having multiple domains implies that there are multiple interfaces, which we can express
\be
\Gamma = \bigcup_{\langle l,m \rangle} \Gamma^{(l,m)} \,,
\label{gamma lm}
\ee 
where we denote by $\Gamma^{(l,m)}$ the interface separating $\Sigma_+^{(l)}$ from $\Sigma_-^{(m)}$, and $\langle l,m\rangle$ is the span over domains sharing an interface. We use $N_\Gamma$ to denote the total number of simple interfaces: for genus zero surfaces it is fixed by the total number of domains as 
\be
N_\Gamma = N_+ + N_- - 1 \,.
\label{genusZeroboundaries}
\ee
It is possible to generalize this relation to higher genera, even if it is not of much use in our context. In fact, \Eqref{genusZeroboundaries} would contain terms depending on both $g$ and on the number of non-contractible interfaces.

The free energy \Eqref{freeenergy} depends on the interface configuration which, as we now explicitly showed, contains both discrete and continuous degrees of freedom. The variational approach adopted in the previous section made the strong assumption that the normal deformation \Eqref{normal deformation} was small enough not to change the domain topology. For this reason, when searching for stable configurations of $F=F[N_+,N_-;\Gamma]$, one should first fix the domain topology, and only then, within a given topological class, look for interface positions which minimize the energy.

If we want to make explicit the sum over connected components, we see that from \Eqref{gamma lm} we should rewrite the line tension term of \Eqref{freeenergy} as
\be
\sum_{\langle l,m \rangle} \sigma \int_{\Gamma^{(l,m)}} \D s \,,
\label{line lm}
\ee
and the terms of \Eqref{freeenergy} involving area integrals should be rewritten as 
\be
\sum_{\alpha=\pm} \sum_{l=1}^{N_\alpha} 
\int_{\Sigma_\alpha^{(l)}} d A 
\left( 
\lambda^{(l)}_\alpha 
+ 
k_\alpha H^2 
+
\bar{k}_\alpha K 
\right) \,,
\label{area lm}
\ee 
where $\lambda_\alpha^{(l)}$ is the Lagrange multiplier relative to the domain $\Sigma_\alpha^{(l)}$. The two constraint equations take the form
\be
\sum_{l=1}^{N_\alpha} \frac{\partial F}{\partial \lambda_\alpha^{(l)}} 
=
\varphi_\alpha A_\Sigma \,,
\label{constraint lm}
\ee
where $\varphi_\pm$ have been introduced in \Eqref{area fractions}. Then, for a variation $\delta \Gamma$ which does not change the topology of domains, \Eqref{interfaceq} becomes
\be
\sigma \kappa_g^{(l,m)}
=
\Delta k \left. H^2 \right|_{\Gamma^{(l,m)}}
+
\Delta \bar{k} \left. K	 \right|_{\Gamma^{(l,m)}}
+
\Delta \lambda^{(l,m)}
\,,
\label{interfaceq lm}
\ee
where $\Delta \lambda^{(l,m)}=\lambda_+^{(l)}-\lambda_-^{(m)}$ and $\kappa_g^{(l,m)}$ is the geodesic curvature of the simple curve $\Gamma^{(l,m)}$. We thus see that for $\Delta k = \Delta \bar{k}=0$, as in \Eqref{CGC equation}, the local minima of \Eqref{freeenergy} consist of CGC curves, each of which with arbitrary curvature, as long as \Eqref{constraint lm} is satisfied.

\section{Curves on minimal surfaces}

\label{app:curves on minimal surfaces}

If the surface is minimal, i.e. it satisfies $H=0$ everywhere, then we have from \Eqref{H gamma} that $\Theta = - \kappa_n$. This implies that the value of Gaussian curvature on $\Gamma$ can be written in terms of Darboux frame quantities,
\be
K = - \left(\kappa_n^2 +\tau_g^2 \right) \,,
\label{minimal K}
\ee
which make manifest how $K\leq 0$ on such surfaces.
By means of the Codazzi-Mainardi equations \Eqref{codazzi mainardi}, we can also express the tangent-normal variation of the Gaussian curvature using only tangential derivatives, finding
\bea
\nabla_{\bm{N}} K &=& 2 \left( \dot{\kappa}_n \tau_g  - \kappa_n \dot{\tau}_g - 2 \kappa_g K \right) \,.
\label{nablaNKmin}
\eea
In fact, this expression can be directly computed from \Eqref{minimal K} using Eqs. \eqref{var kn taug}. For an alternative derivation, see Appendix A of \cite{envirobias}.

If we switch to the frame of principal directions, where we define $\theta$ to be the angle along $\Gamma$ between $\bm{T}$ and $\bm{k}_1$, the minimality conditions allows several further simplifications. Namely the geodesic curvatures of the principal directions can be written as derivatives of the Gaussian curvature (see Theorem 3.3 of Ref. \cite{ando2011geodesic})
\be
\kappa_g(\bm{k}_i) = \frac{1}{4K}\,\epsilon_{i}^{j} \partial_j K   \,, 
\ee
where $\epsilon_{i}^{j}=h^{ik}\epsilon_{jk}$, with $\epsilon_{12}=-\epsilon_{21}=\sqrt{h}$, is the two-dimensional Levi-Civita tensor. This implies that the geodesic curvature of an arbitrary curve on a minimal surface can always be written as
\be 
\kappa_g = \dot{\theta} - \frac{1}{4 K} \nabla_{\bm{N}} K  \,.
\label{kappagthetamin}
\ee
By substituting the definitions \Eqref{kappantheta} and \Eqref{taugtheta} into \Eqref{nablaNKmin}, we re-obtain \Eqref{kappagthetamin}, confirming the consistence of these results.
Furthermore, the geodesic torsion and normal curvature are related to $\theta$ via 
\be
\tan 2 \theta = \frac{\tau_g}{\kappa_n} \,. 
\label{tanthetamin}
\ee
In the principal directions frame we can compute explicitly the on-shell second variation \Eqref{var var F}, finding
\be 
\sigma 
\int_\Gamma \D s \left\{
(\dot{\epsilon}_{\bm{N}})^2
-
\epsilon_{\bm{N}}^2 
\left[
\kappa_g^2
+
K 
\left(
1-4 \etakb (\kappa_g - \dot{\theta})
\right)
\right]
\right\} \,.
\label{var var F min}
\ee
If $\varphi$ is not conserved, we have that $\Gamma$ obeys $\kappa_g = \etakb K$: such an interface must then satisfy the stability condition
\be
1+ 4 \etakb \dot{\theta} + 3 \etakb^2 (\kappa_n^2+\tau_g^2) \geq 0\,.
\ee
Since only the second term is not necessarily positive, this inequality implies that the interface cannot deviate too quickly from a principal direction.

\subsection{The Weierstrass representation}

\label{app:weierstrass}

Every simply connected minimal surface can be described using its Weierstrass-Enneper representation \cite{Colding2011}. This is an explicit parametrization of the form
\be
\bm{r}(u,v) = \re e^{i \theta_B} \int_{\omega_0}^\omega dz \, \bm{\phi}(z) \,,
\label{WE representation}
\ee
where $\omega = u + i v$ is a complex number and $\omega_0=u_0+i v_0$ is such that $x^\mu(u_0,v_0)$ belongs to the surface. The phase $\theta_B$ is known as the Bonnet angle. The vector field $\bm{\phi}: \mathbb{C} \to \mathbb{C}^3$ has the crucial property
\be
\bm{\phi} \cdot \bm{\phi} = \left( |\re\bm{\phi}\,|^2 - |\im\bm{\phi}\,|^2 \right) + 2 i \im \bm{\phi}\, \cdot \re \bm{\phi} = 0 \,.
\label{Phi squared}
\ee
In fact, this condition guarantees that the coordinates $\{u,v\}$ are isothermal: since the tangent vectors are
\be
\bm{t}_u = \re  e^{i \theta_B}\bm{\phi} \,,\quad 
\bm{t}_v = -\im  e^{i \theta_B}\bm{\phi}  \,,
\label{WE tangent}
\ee
they are mutually orthogonal and have identical norm, so that the induced metric \Eqref{induced metric} is conformally flat
\be
h_{ij} = \Omega^2 \delta_{ij} \,,
\label{hij conf flat}
\ee
with $\Omega=\Omega(u,v)$ being the conformal factor.
Although in two dimensions every surface $\Sigma$ can be locally parametrized with isothermal coordinates, i.e. it is possible to find a pair of coordinates such that the induced metric takes the form \Eqref{hij conf flat}, this procedure is useful practically only for minimal surfaces. In fact, the surface parametrized by \Eqref{WE representation} is minimal if and only if the mapping $\bm{r}(u,v)$ is harmonic, i.e. it satisfies $(\partial_u^2 +\partial_v^2)\bm{r}=0$. In turn, this is true if and only if the function $\bm{\phi}$ is analytic. Non-minimal surfaces would have a non-analytical $\bm{\phi}$, rendering the integral representation \Eqref{WE representation} not particularly illuminating. 

Since $\bm{\phi}$ has three components but must satisfy \Eqref{Phi squared}, every minimal surface is then completely determined by two complex analytic functions. A possible explicit choice for the parametrization is
\be
\bm{\phi}(z) = \left(
\begin{array}{c}
f (1-g^2) \\
i f(1+g^2) \\
2 f g
\end{array}
\right)  \,,
\label{WE rep 1}
\ee
where $f=f(z)$ is analytic and $g=g(z)$ is meromorphic but such that $f g^2$ is analytic. These functions are defined over a suitable domain $\mathcal{D}\in \mathbb{C}$. 

The Bonnet angle is a free parameter which defines a family of surfaces with identical intrinsic geometry: the induced metric \Eqref{hij conf flat} does not depend on the value of $\theta_B$: all surfaces belonging to the same family are thus locally isometric. 

Given \Eqref{WE rep 1}, the conformal factor is $\Omega=|f|(1+|g|^2)$. We can then compute the extrinsic curvature tensor
\be
K_{ij} = 2 \left(
\begin{array}{c c}
-\re e^{i \theta_B} f g_z & \im e^{i \theta_B} f g_z \\
\im e^{i \theta_B} f g_z & \re e^{i \theta_B} f g_z
\end{array}
\right)  \,,
\label{WE Kij}
\ee
with the shorthand notation $g_z=\partial_z g$. \Eqref{WE Kij} immediately shows how the minimality condition, $H=0$, is satisfied. From this, the Gaussian curvature is readily obtained
\be
K = -\frac{4 |f g_z|^2}{\Omega^4} \,.
\label{WE K}
\ee
As expected, it is always non-positive and does not depend on the Bonnet angle. Note that poles of $f$ become zeroes of the Gaussian curvature.

We can now use the geometrical tools defined in Appendix \ref{app:curves} to understand how curves behave on minimal surfaces. The interface $\Gamma$ is defined on $\mathcal{D}$ via the complex one-parameter curve $z(s)=u(s)+i v(s)$. The unit tangent and normal to $\Gamma$ are thus
\begin{subequations}
\begin{align}
\bm{T}&= \cos\alpha \, \bm{t}_u + \sin \alpha \, \bm{t}_v \,,\\
\bm{N}&= \sin\alpha \, \bm{t}_u - \cos \alpha \, \bm{t}_v \,,
\end{align}
\end{subequations}
where we defined $\alpha$ to be the angle between $\bm{T}$ and $\bm{t}_u$. It is then immediate to see that $z'(s)= \frac{1}{\Omega} e^{i \alpha}$. We chose the convention on the orientation of the rotation such that if $\bm{T}$ is pointing in the direction $\alpha=0$ at the origin, then the $"+"$ domain lies in the $\im z < 0$ part of the complex plane.

Since $\partial_z = \frac{1}{2} (\partial_u - i \partial_v)$, for any real scalar function $A(u,v)=A(z,\bar{z})$ we have that both tangential and normal variations can be expressed in terms of the real and imaginary parts of derivatives with respect to $z$
\begin{subequations}
\begin{align}
\frac{\D A}{\D s}
=
\nabla_{\bm{T}} A 
&=
 \frac{2}{\Omega} \re e^{i\alpha} \partial_z A\,, 
\\
\nabla_{\bm{N}} A 
&=
 \frac{2}{\Omega} \im e^{i\alpha}  \partial_z A \,.
\end{align}
\end{subequations}
Interestingly, we find a particularly compact expression for the geodesic curvature of $\Gamma$
\bea
\kappa_g 
&=& \dot{\alpha} + \nabla_{\bm{N}} \log \Omega \nn \\
&=& \dot{\alpha} + \frac{2}{\Omega} \im e^{i \alpha} \partial_z \log \Omega \,.
\label{kappag WE}
\eea
Similarly, from \Eqref{WE Kij} we can compute the normal curvature and the geodesic torsion
\begin{subequations}
\begin{align}
\kappa_n &= -\frac{2}{\Omega^2} \re e^{2i\alpha} f g_z \,,
\label{kappan WE}
\\
\tau_g &= -\frac{2}{\Omega^2} \im e^{2i\alpha} f g_z \,,
\label{taug WE}
\end{align}
\end{subequations}
which indeed satisfy the relation $K=-(\kappa_n^2+\tau_g^2)$. Note that $\alpha$ is related to $\theta$ via \Eqref{tanthetamin}:
\be
\tan 2 \theta = \frac{\im e^{2i\alpha} f g_z}{\re e^{2i\alpha} f g_z} \,.
\ee
This shows how, if $f g_z$ is constant, that  $u$ and $v$ coordinates are rotated by the Bonnet angle with respect to the principal directions. In general, however, there is no simple relationship between the isothermal coordinates and the principal directions.
Furthermore, since we have
\be
 \dot{\kappa}_n \tau_g - \kappa_n \dot{\tau}_g =
 K \left( 2 \dot{\alpha} +\im  \frac{e^{i \alpha}}{\Omega} \partial_z \log f g_z \right) \,,
\label{kn taug WE}
\ee
we can compute explicitly the normal variation of the Gaussian curvature:
\be
\nabla_{\bm{N}}K 
= 2\,K\; \im \frac{e^{i \alpha}}{\Omega} \partial_z \log \frac{f g_z}{\Omega^4} \,.
\label{nablaN K WE}
\ee
By combining together \Eqref{kappag WE}, \Eqref{kn taug WE} and \Eqref{nablaN K WE} one can see that \Eqref{nablaNKmin} is indeed satisfied.

\subsection{The Schwarz P-surface and its nodal approximation}
\label{app:schwarz P}

An arbitrary pair of analytic functions $f(z)$ and $g(z)$ produces always a minimal surface, which however is not necessary an embedding. Nonetheless, in the literature several families of embedded minimal surfaces are known. In fact, if we choose the functions $f(z)=\frac{1}{\sqrt{z^8- 14 z^4 +1}}$, $g(z)=z$ and fix the Bonnet angle to be $\theta_B=\pi/2$ we obtain the Schwarz P-surface. 

The complex variable $z$ of \Eqref{WE rep 1} lies in the \textit{fundamental patch} $\mathcal{D}=\mathcal{D}_+ \cup \mathcal{D}_-$, where
\be
\mathcal{D}_\pm = \{ z \in \mathbb{C} \,: \im z \geq 0,\, \re z \geq 0,\, |z-c_\pm|^2 \leq 2 \}  \,,
\label{WE D}
\ee
with $c_\pm=-\frac{1}{\sqrt{2}}(1\pm i)$. The analytic expression of the explicit parametrization can be found e.g. in \cite{Gandy2000}, and is
\bea
\bm{r}(u,v) &=& 
\frac{\mu}{4} 
\left\{
\begin{array}{c}
\sqrt{2} \im  F\left( p,\frac{1}{4}\right)\\
\sqrt{2} \re  F\left(p,\frac{1}{4}\right) \\
\im F\left(q, 97 -56 \sqrt{3}\right) 
\end{array}
\right\} \,,
\eea
with $p=\arcsin	\frac{\sqrt{2}\omega}{\sqrt{\omega^4+4\omega^2+1}}$, $q=\arcsin\frac{4 \omega^2}{\omega^4+1}$ and $\omega=u+iv$. $F$ is the incomplete elliptic integral of the first kind (with the notation convention used by \textit{Mathematica}'s function \textit{EllipticF}).

Incidentally, the Schwarz D and the \textit{Gyroid} surfaces belong to the same Bonnet family of the P-surface, having respectively $\theta_B=0$ and $\theta_B=K(3/4)/K(1/4)$.

The specification of the WE functions is enough to compute the interface shape and its stability. The explicit parametrization serves only as a mapping to $\mathbb{R}^3$, and is useful for graphical representations (see right panel in \fref{fig:figure schwarz}).

For all practical purposes, however, one needs to reconstruct the full surface past the boundaries of the fundamental patch. This is done by using the so called Schwarz reflection principle (see e.g. \cite{Colding2011}), which allows to extend the surface by means of a specific subgroup of the Euclidean isometries. This subgroup eventually determines the crystallographic group of the surface, which for the Schwarz P surface is $Pm\bar{3}m$ - hence the name. The unit cell shown on the left of \fref{fig:figure schwarz} consists of $48$ fundamental patches glued together.

Even if the extension of the surface via reflections, translation and rotations is quite straightforward, this is not the case for embedded curves. In fact, when crossing the boundary between two different patches, one needs to keep track of both position and the full co-moving Darboux frame: this is certainly possible, but rather laborious.

We choose instead a different path to overcome this difficulty: we make use of the \textit{nodal approximation} of the surface \cite{VonSchnering1991}. This is obtained by truncating the Fourier series expansion of a field whose zero level set define the minimal surface \cite{Schwarz1999}. The first-order truncation leads to the implicit relation
\be
\cos \frac{2 \pi x}{L} + \cos \frac{2 \pi  y}{L} + \cos \frac{2 \pi z}{L} = 0 \,,
\label{nodal implicit}
\ee
with $L$ the width of a unit cell.
Although non minimal, the space group of the nodal surface is identical to the one of the P surface. Higher orders terms in the expansion involve more complicated combinations of trigonometric functions \cite{Gandy2001}. 

The advantage of \Eqref{nodal implicit} is that we can express the surface as the union of an infinite stack of Monge patches of the form
\be 
z_{\pm,n}(x,y) = L \left[ n\pm \arccos \left(-\cos \frac{2 \pi  y}{L}-\cos \frac{2 \pi  y}{L} \right)  \right]
\label{nodal layer}
 \,,
\ee
with $n \in \mathbb{N}$. This function is defined for all $x,y$ provided the argument of the inverse cosine is between $-1$ and $1$. In this way, numerical solutions of the interface equation can be easily obtained for big portions of the surface without having to worry about gluing conditions. In \fref{fig:figure nodal} we show the patches $z_{+,0}$ and $z_{-,1}$.

\section{Curves on developable surfaces}

\label{app:curves on developable surfaces}

A developable surface has by definition $K=0$, implying $\Theta = \tau_g^2 / \kappa_n$. Since both principal directions are geodesics, we have $\kappa_g=\dot{\theta}$. The  mean curvature is then
\be
H= \frac{\kappa_n^2+\tau_g^2}{\kappa_n} \,, 
\ee
unless $\kappa_n=0$, for which $H=0$. Using the Codazzi-Mainardi equations \eqref{codazzi mainardi}, we find that the normal variation of the squared mean curvature can be expressed only in terms of tangential derivatives 
\bea
\nabla_{\bm{N}} H^2 
&=&  \tan \theta \; \frac{\D}{\D s} H^2 
\,.	
\eea
Because of $H^2>0$, every point where $H=0$ is necessarily a minimum for $H^2$. Furthermore, developable surfaces satisfy the relation
\be
\tan \theta = \frac{\tau_g}{\kappa_n} \,,
\label{tanthetadev}
\ee
with a factor of two difference with respect to \Eqref{tanthetamin}. The full on-shell second variation reduces to
\be 
\delta^{(2)} F
=
\sigma 
\int_\Gamma \D s \left[
(\dot{\epsilon}_{\bm{N}})^2
-
\epsilon_{\bm{N}}^2 
\left(
\dot{\theta}^2 
+
\ddot{\theta} \tan \theta
\right)
\right] \,,
\label{var var F dev}
\ee
which, for non-fixed $\varphi$, leads to \Eqref{stability dev}.

The profile and the curvature are always one-parameter functions. For instance, consider a cylindrical developable surface, let $x$ to be the direction where the curvature varies and let $y$ be the flat direction so that we can parametrize the surface with a single function $z=h(x)$. The mean curvature is then
\be
H= \frac{h''(x)}{2(1+h'(x)^2)^{3/2}}  \,.
\label{H dev monge}
\ee

\subsection{Closed orbits in a varying magnetic field}

\label{app:orbits in a B}

In this Section we show that finding interfaces on developable surfaces is mathematically equivalent to solving the equation of motion for a charged particle in a spatially varying magnetic field.  

Any interface on $\Sigma$ must be a solution of the ODE system comprising of \Eqref{interface dev} and of the arc-length condition. We parametrize the unit tangent vector as
\be
\bm{T} = \cos \theta \bm{k}_1 + \sin \theta \bm{k}_2 \,,
\label{t dev 0}
\ee
where $\bm{k}_1$ is the curved principal direction. The embedding functions $\{x(s),y(s)\}$ are linked to the angle $\theta$ via
\begin{subequations}
\begin{align}
\dot{x} &= \bm{T} \cdot \bm{k}_2 = \cos\theta \,,
\label{t dev 1}\\
\dot{y} &= \bm{T} \cdot \bm{k}_2 = \sin\theta \,.
\label{t dev 2}
\end{align}
\label{t dev}
\end{subequations}
We stress that $x$ and $y$ are the two parametric coordinates of the surface (in Appendix \ref{app:surfaces} were called $\sigma^1$ and $\sigma^2$). They coincide with standard Cartesian coordinates only for developable surfaces which are invariant under rigid translations, as the wave-like profile of \fref{fig:figure developable 1}. However, note that every result in the following does not require $\Sigma$ to be cylindrical.
A unique solution of Eqs. \eqref{interface dev} and \eqref{t dev} is fixed by the choice of three initial conditions $x(0)=x_0$, $y(0)=y_0$ and $\theta(0)=\theta_0$.

By replacing the principal directions in \Eqref{t dev 0} with the standard Cartesian axes $\bm{k}_1 \to \bm{\hat{x}}$ and $\bm{k}_2 \to \bm{\hat{y}}$, the curve $\Gamma$ becomes a curve in $\mathbb{R}^2$. This new curve has tangent and normal vectors
\begin{subequations}
\begin{align}
\bm{v} &= \cos \theta \bm{\hat{x}} + \sin \theta \bm{\hat{y}} \,,
\label{v dev}\\
\bm{\tilde{N}} &= -\sin \theta \bm{\hat{x}} + \cos \theta \bm{\hat{y}} \,.
\label{N tilde dev}
\end{align}
\label{vN dev}
\end{subequations}
We choose to interpret $\bm{v}$ as the \textit{velocity} of a particle moving along a trajectory on the $xy$ plane, with the arc-length $s$ playing the role of time. At $s=0$ the particle is passing through the point $(x_0,y_0)$ with unit velocity in the direction $\theta_0$. The geodesic curvature is then naturally interpreted as the instantaneous \textit{centripetal acceleration} of the particle
\be
\kappa_g = \bm{\tilde{N}} \cdot \dot{\bm{v}}  \,.
\ee
Since the speed is constant, $\bm{v}\cdot\bm{v}=1$, there is no tangential component of $\dot{\bm{v}}$.

The right-hand side of \Eqref{interface dev} is interpreted as an induced normal acceleration:  this is exactly the type of force that a charged particle would experience while moving in a spatially varying magnetic field. In fact, let us define
\be
\bm{B} = B_z \bm{\hat{z}} \,,
\ee
with $B_z(x)= \etak H(x)^2 + \frac{\Delta \lambda}{\sigma}$ (see \Eqref{H dev monge}). Then \Eqref{interface dev} is equivalent to 
\be
\dot{\bm{v}} = \bm{v} \times \bm{B} \,,
\label{Lorentz force}
\ee
which clearly shows how the right-hand side is essentially the Lorentz force. To see the equivalence, it is sufficient to take the contraction of \eqref{Lorentz force} with the unit in-plane normal $\bm{\tilde{N}}$. Since the velocity is orthogonal to the magnetic field, i.e. $\bm{v}\cdot \bm{B}=0$, the particle will follow a planar trajectory.

We can now trade intuition from electromagnetism to get insight into the interface problem. The orbits should be closed and simple. This is always the case for a constant non-vanishing magnetic field, corresponding to either a flat surface or to $\etak=0$, and the orbit radius is given by $\sigma/\Delta \lambda$ (we formally set the mass and the charge of the particle to one).  For non-conserved $\varphi$ the radius diverges and the trajectory is a straight line, i.e. the path followed by free particle.

If $\bm{B}$ is varying with $x$, then the closeness of $\Gamma$ becomes a less trivial requirement. In fact, for a given $H(x)$ and specific initial conditions, there usually are only a \textit{discrete} set of values $\Delta\lambda$'s which allow for closed orbits. In general the particle will have a trajectory of non-constant curvature and will drift towards a direction perpendicular to both the magnetic field and its gradient. In our case, the particle will drift along $\bm{\hat{y}}$. The center of the instantaneous osculating circle of the orbit is known as the \textit{guiding center}.

All this can be easily proven within our framework. First, note that the free energy \Eqref{freeenergy} is invariant under translations along the $y$ direction. The analogy with the magnetic field tells us immediately that there is a conserved charge: the \textit{minimally coupled momentum}.

We can always find an electromagnetic potential $\bm{A}$ such that $\bm{B}= \nabla \times \bm{A}$. We fix the gauge so $\bm{A}$ has only one non-zero component:
\be
\bm{A} = A_y \bm{\hat{y}} \,.
\ee
In classical mechanics, the coupling between a charged particle and an electromagnetic field is captured by the substitution
\be 
\bm{v} \to \bm{P} = \bm{v} - \bm{A}  \,,
\label{minimal momentum}
\ee
in the Lagrangian. The $y$-component of the momentum $\bm{P}$ is conserved along the orbit:
\be 
\dot{P}_y = 0 \,,
\label{conserved Py}
\ee
which can be directly proven by multiplying \Eqref{interface dev} with \Eqref{t dev 1} and using $B_z = \partial_x A_y$. 

The condition that the interface has to be a simple closed curve implies that its total length $\ell_\Gamma$ is determined via the Gauss-Bonnet theorem \Eqref{GB theo kg} as
\be 
\int_\Gamma \D s  \kappa_g = \theta(\ell_\Gamma)- \theta_0 = 2 \pi \,,
\label{GB dev}
\ee
i.e. $\ell_\Gamma$ is the ``time'' after which $\bm{v}$ points in the same direction as it was pointing at $s=0$. For a closed orbit, this time corresponds to the orbital period. 

If $\theta$ is a periodic function, then also $\dot{\theta}$ is periodic. Because $\dot{\theta}(s)=B_z(x(s))$, this implies that $x(s)$ has to be periodic in $s$ as well, and the average displacement along the $x$ direction vanishes 
\be
\Delta x = \left< v_x \right>_\Gamma = \int_\Gamma \D s \dot{x} = 0 \,,
\ee
where we introduced the notation $\left< \cdot \right>_\Gamma$ for averages over a single orbit.
Conversely, even if $\dot{y}$ is periodic we cannot conclude that $y$ will be periodic. In fact, in general $\Delta y =  \left< v_y \right>_\Gamma \neq 0$ and the particle will drift along the flat direction. By tuning the value of $\Delta \lambda$ - i.e. changing the average value of the magnetic field - it is possible to find orbits that have $\Delta y =0$. 

If $\Sigma$ has either some reflection planes or inversion lines along the $y$-axis, then interfaces crossing orthogonally the symmetry line will always be closed, for \textit{any} $\Delta \lambda$. In fact, this is true also for rotationally invariant surfaces, and is the reason why every interface - besides geodesics - of \fref{fig:tube} is closed.

The analogy with the magnetic field is insightful even for the energy functional: the area integral of \Eqref{freeenergy} is the \textit{magentic flux} $\Phi_{\bm{B}}$ of the field $\bm{B}$ through the flat, compact surface $\Sigma_\pm$
\be
\Phi_{\bm{B}}(\Sigma_\pm) = \int_{\Sigma_\pm} \D A B_z  \,,
\ee 
where the domain of integration is bounded by the particle's orbit, $\partial \Sigma_\pm = \Gamma$. 

We can reduce area integrals to line integrals: using Stokes' theorem we can compute the area fraction occupied by a soft domain
\bea
\varphi &=&\frac{1}{2 A_\Sigma} \left< \bm{\tilde{N}} \cdot \left(x \bm{\hat{x}}+y \bm{\hat{y}} \right) \right>_\Gamma \nn 
\,,
\eea
where $\bm{\tilde{N}}$ is the in-plane normal to $\bm{v}$ (see \fref{fig:figure developable 1}) and $x \bm{\hat{x}}+y \bm{\hat{y}}$ is the particle position. Similarly, the magnetic flux can be written as the circulation of the electromagnetic potential
\be 
\Phi_{\bm{B}}(\Sigma_\pm) = \mp \left< \bm{v} \cdot \bm{A} \right>_\Gamma =  \mp \left< v_y^2 \right>_\Gamma  \,,
\ee
where in the second identity we used \Eqref{conserved Py} and the fact that $\Delta y=0$ on closed orbits.
Since $\Delta k = k_+ - k_-$, the magnetic flux entering in the free energy is the one relative to the $\Sigma_+$ domains. We finally find the compact expression
\be
F = \sigma \ell_\Gamma + \sigma \Phi_{\bm{B}}(\Sigma_+) = \sigma \left< v_x^2 \right>_\Gamma \,,
\label{freeenergy dev 3}
\ee
i.e. the $F$ is proportional to the one-orbit average kinetic energy in the $x-$direction. 

Because $A_y$ is defined up to an arbitrary integration constant, we can fix the condition $\left< A_y \right>_\Gamma=0$. In turn, this implies $P_y=0$. With this choice, \Eqref{freeenergy dev 3} becomes
\be
F = \sigma \int_\Gamma \D s  \left(\bm{v}-\bm{A} \right) \cdot \left(\bm{v}-\bm{A} \right) \,,
\ee
which is precisely the Lagrangian of a minimally coupled charged particle.

\bibliography{Leiblery}{}

\end{document}